\documentclass[10pt,journal,compsoc]{IEEEtran}

\usepackage{cite}

\usepackage{amsfonts}
\usepackage[caption=false,font=footnotesize,labelfont=sf,textfont=sf]{subfig}
\usepackage{multirow}
\usepackage{xcolor}
\usepackage{mathrsfs}
\usepackage{amsmath}
\usepackage{amsthm}
\usepackage{url}
\usepackage{graphicx}
\usepackage{booktabs}
\usepackage{amssymb}
\usepackage{balance}
\usepackage[citecolor=blue, colorlinks]{hyperref} 
\usepackage{makecell}
\begin{document}

\title{Light Field Reconstruction via Deep Adaptive Fusion of Hybrid Lenses}

\author{Jing Jin,
        Mantang Guo,
        Junhui Hou,~\IEEEmembership{Senior Member, IEEE},
        Hui Liu,
        and Hongkai Xiong,~\IEEEmembership{Senior Member, IEEE }
        \thanks{This work was supported in part by the Hong Kong Research Grants Council under Grants 11218121 and 21211518, in part by the Hong Kong Innovation and Technology Fund under Grant MHP/117/21, in part by the Basic Research General Program of Shenzhen Municipality under Grant JCYJ20190808183003968, and in part by Hong Kong University Grants Committee under Grant UGC/FDS11/E02/22. (\textit{Corresponding author: Junhui Hou})
        }
\thanks{J. Jin, M. Guo, and J. Hou are with the Department of Computer Science, City University of Hong Kong, Hong Kong, and also with the City University of
Hong Kong Shenzhen Research Institute, Shenzhen 518057, China (e-mail: jingjin25-c@my.cityu.edu.hk; mantanguo2-c@my.cityu.edu.hk; jh.hou@cityu.edu.hk).}
\thanks{H. Liu is with the School of Computing and Information Sciences, Caritas
Institute of Higher Education, Hong Kong. (e-mail:hliu99-c@my.cityu.edu.hk;) }
\thanks{H. Xiong is with the Department of Electronic
Engineering, Shanghai Jiao Tong University, Shanghai 200240, China (e-mail: xionghongkai@sjtu.edu.cn).}

}

\markboth{MANUSCRIPT SUBMITTED TO IEEE TRANSACTIONS ON PATTERN ANALYSIS AND MACHINE INTELLIGENCE}
{Shell \MakeLowercase{\textit{et al.}}: Bare Demo of IEEEtran.cls for IEEE Journals}

\IEEEtitleabstractindextext{
\begin{abstract}
    This paper explores the problem of reconstructing high-resolution light field (LF) images from hybrid lenses, including a high-resolution camera surrounded by multiple low-resolution cameras. The performance of existing methods is still limited, as they produce either blurry results on plain textured areas or distortions around depth discontinuous boundaries. To tackle this challenge, we propose a novel end-to-end learning-based approach, which can comprehensively utilize the specific characteristics of the input from two complementary and parallel perspectives. Specifically, one module regresses a spatially consistent intermediate estimation by learning a deep multidimensional and cross-domain feature representation, while the other module warps another intermediate estimation, which maintains the high-frequency textures, by propagating the information of the high-resolution view. We finally  leverage the advantages of the two intermediate estimations adaptively via the learned confidence maps, leading to the final high-resolution LF image with satisfactory results on both plain textured areas and depth discontinuous boundaries. Besides, to promote the effectiveness of our method trained with simulated hybrid data on real hybrid data captured by a hybrid LF imaging system, we carefully design the network architecture and the training strategy. Extensive experiments on both real and simulated hybrid data demonstrate the significant superiority of our approach over state-of-the-art ones. To the best of our knowledge, this is the first end-to-end deep learning method for LF reconstruction from a real hybrid input.  We believe our framework could potentially decrease the cost of high-resolution LF data acquisition and benefit LF data storage and transmission. The code will be publicly available at \url{https://github.com/jingjin25/LFhybridSR-Fusion}.
\end{abstract}

\begin{IEEEkeywords}
Light field, super-resolution, hybrid imaging system, deep learning, fusion, depth.
\end{IEEEkeywords}
}
\maketitle

\IEEEdisplaynontitleabstractindextext
\IEEEpeerreviewmaketitle

\IEEEraisesectionheading{\section{Introduction}\label{sec:introduction}}

\IEEEPARstart{T}{he} light field (LF) describes all light rays through every point along every direction in a free space \cite{levoy1996lf}.
An LF image can be interpreted as multiple views observed from viewpoints regularly distributed over a 2-D grid. Therefore, LF images contain not only color information but also geometric structure of the scene in an implicit manner. The rich information enables many applications such as 3-D reconstruction \cite{lfapp2013scene}, image post-refocusing \cite{lfapp2014refocusing}, material recognition \cite{lfapp2016material}, saliency detection \cite{lfapp2014saliency},  densely-sampled LF reconstruction \cite{guo2021learning}, and motion deblurring \cite{lfapp2017motion}.  
Recent research also demonstrates that LF is a promising media for virtual/augment reality \cite{lfapp2015vr,lfapp2017vryu}.

\begin{figure*}[t]
    \begin{center}
    \includegraphics[width=\linewidth]{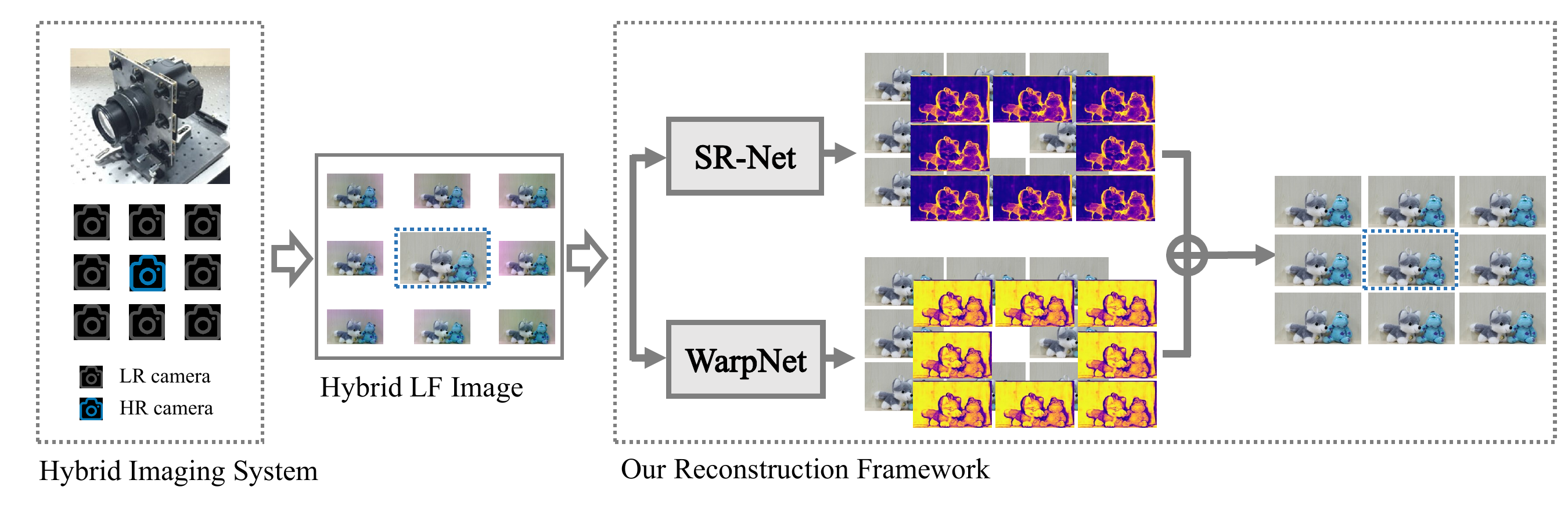}
    \end{center}
    \vspace{-1.0em}
      \caption{Illustration of the proposed framework. The hybrid imaging system \cite{lfhybrid2017ring} captures an HR central view and multiple LR side views. Two sub-networks that are complementary to each other are involved to reconstruct the HR LF image, and the predictions of them are adaptively fused based on learned confidence maps. 
      Specifically,  the SR-Net spatially super-resolves the input LR side views under the guidance of the HR central view, and the Warp-Net warps the HR central view with the disparity maps estimated from the LR side views. Finally, the predictions produced by these two modules are adaptively fused based on the learned confidence maps to generate an HR LF image.  The blue frames indicate that the central view of the reconstructed HR LF comes from the input.  
      }
    \label{fig:flowchat}
    \end{figure*}

A high-quality LF image can be captured by a densely positioned array of high-resolution (HR) cameras. However, it is neither practical nor necessary to do so with so many separate HR units.
Recent commercialized  LF cameras provide a convenient way to capture LF images. However, the captured LF images always suffer from low spatial resolution due to the limitation of sensor resolution. 
To overcome this limitation, many methods for reconstructing HR LF images have been proposed~\cite{lfssr2014variational,lfssr2017graph,lfssr2015yoon,lfasr2020aaai,lfasr2020eccv,lfasr2018Yeung,lfasr2020pami,lfhybrid2018dsp,lfhybrid2014iccp,lfhybrid2017ring,lfhybrid2016splitter,lfhybrid2018tci}.
Among them, LF reconstruction with a hybrid input is a promising way. 
A hybrid LF imaging system can be built by a sparse grid of low-resolution (LR) image sensors that surround a central HR camera ~\cite{lfhybrid2017ring,lfhybrid2018tci},  as shown in Fig. \ref{fig:flowchat}.
These heterogeneous sensors simultaneously sample along the angular and spatial dimensions of the LF at different sampling rates, and provide sufficient information for subsequent algorithms to calculate an HR LF.
The LR views are useful for recording the geometry information of the scene, while the HR central view captures delicate textures and high-frequency information of the scene. To produce an HR LF image, a post-process algorithm is necessary to combine the information of the hybrid input.

Although multiple algorithms have been proposed to reconstruct an HR LF from the hybrid input \cite{lfhybrid2014iccp,lfhybrid2016splitter,lfhybrid2018tci,lfhybrid2017ring}, they still have limited performance. Generally, these methods comprise several steps that are independently designed, and the final results would be compromised by any inaccuracy of each step. Furthermore, these methods fail to fully describe the complicated relation between 
the HR central view and the LR side views as well as the one within the high-dimensional LF image.

We propose a learning-based framework to reconstruct an HR LF image with a hybrid input in an end-to-end manner. 
The proposed framework produces impressive performance. As illustrated in Figure \ref{fig:flowchat}, our framework achieves the goal with two \textit{complementary} and \textit{parallel} research lines, namely SR-Net and Warp-Net, and the advantages of them are combined via confidence-guided fusion.
The SR-Net up-samples the LR views to the desired resolution by learning a deep representation from both components of the hybrid input. The results of this module are spatially consistent concerning the scene content but always blurred, especially when the up-sampling scale is relatively large. In Warp-Net, the HR view is warped to synthesize an HR LF using the disparity maps estimated from the LR views. The predictions by this module inherit the delicate textures and high-frequency information from the HR view but always have artifacts caused by occlusion or disparity inaccuracy. Observing the complementary behavior between these two modules, we learn a pixel-wise confidence map for the output of each module. And the final HR LF image is obtained by adaptively fusing the two intermediate predictions based on their  confidence maps, in which only their advantages are collected.

This paper follows the overall framework proposed in our previous conference paper \cite{lfhybrid2020fusion}, namely HybridLF-Net. Yet, HybridLF-Net was merely designed for simulated hybrid data, i.e., the LR side views are generated by down-sampling an HR LF image, and its effectiveness on real hybrid data captured from a typical hybrid imaging system is not explicitly considered.
To be specific,
the SR-Net of HybridLF-Net explores the LF features using spatial-angular separable (SAS) convolutions and the Warp-Net of HybridLF-Net  estimates disparity maps using a plain and shallow convolutional network on the LR LF image.
However, there is a significant gap between real and simulated hybrid data, such as the color inconsistency across views, the relatively large disparity, and the inaccurate LF structure among views (i.e., the LR side views and the down-sampled HR central view no longer form an accurately calibrated LF image). Consequently, HybridLF-Net cannot work well on 
real hybrid data. That is, the SAS-based feature extraction manner in SR-Net and the LF-based disparity estimation are inappropriate, and  the accuracy of the disparity maps estimated by Warp-Net is insufficient. See the quantitative and qualitative results in Sec. \ref{sec:exp}.

Being aware of these challenges, we carefully redesign both the network architecture and the training strategy to promote the effectiveness of the framework  
on real hybrid data. 
Particularly, we make the following efforts:
\begin{enumerate}
  \item  we remove 
   the central view from the LR views to avoid the influence of different central view characteristics between training and testing,  
   and accordingly
   we modify the SR-Net and Warp-Net to adapt them to the stack of LR side views instead of the LF image for learning the LF representations and explicit geometry;  
  
  \item  in the Warp-Net, we use a multi-scale structure to explore the long-distance correlations among views;

  \item   we 
  further enhance the utilization of the high-frequency information of the HR central view from two perspectives, i.e., in the SR-Net, the HR features are fused with the features of each side view more sufficiently, and in the Warp-Net, the HR details of the central view are utilized to enhance the estimated disparity maps estimated from the LR views; and

  \item we propose a training strategy tailored to real data, i.e.,  
   LF images with large disparities and augmented with color perturbation are used to construct the training dataset.
\end{enumerate}

Benefiting from the carefully designed training strategy and network architecture, our framework trained with simulated hybrid data can work well on real hybrid data.
Extensive experiments on the hybrid data captured by a real imaging system, as well as that simulated from LFs, demonstrate the significant superiority of our method over HybridLF-Net \cite{lfhybrid2020fusion}, as well as other state-of-the-art ones. 
That is, our method can reconstruct HR LF images with  higher quality and better parallax structure effectively and efficiently.

The rest of this paper is organized as follows. Sec. \ref{sec:rw} comprehensively reviews existing methods for  image super-resolution. Sec. \ref{sec:proposed method} presents the proposed method. Sec. \ref{sec:exp} demonstrates the advantages of the proposed method through extensive experiments on both real and simulated hybrid data.
Finally, Sec. \ref{sec:con} concludes this paper.

    \begin{figure*}
    \begin{center}
    \includegraphics[width=0.95\textwidth]{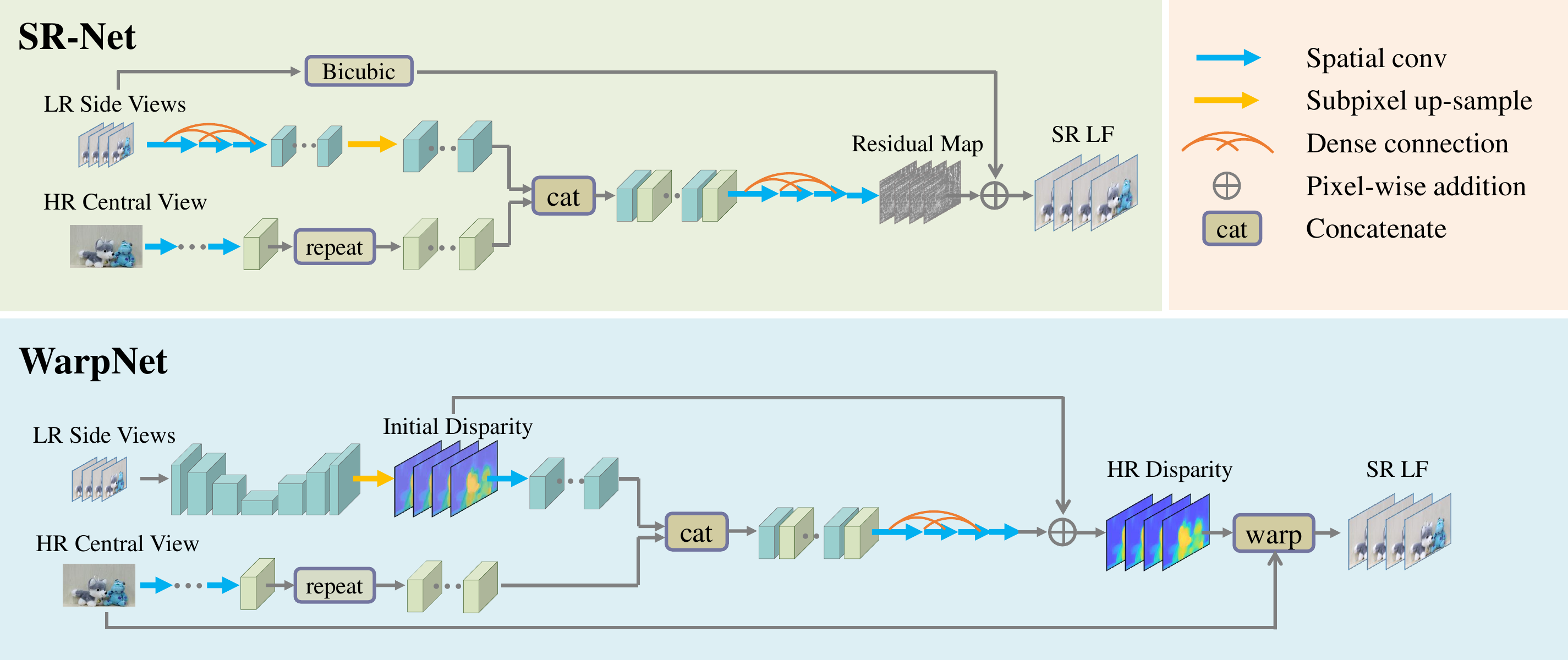}
    \end{center}
    \vspace{-0.5em}
      \caption{Illustration of the network architectures of the SR-Net and the Warp-Net. 
      The SR-Net super-resolves the LR side views by fusing the high-frequency information from the HR central view, while the Warp-Net synthesize an HR LF image by warping the HR central view based on the disparity map estimated from the LR side views. We refer readers to the supplementary file for the detailed architecture.
      }
    \label{fig:subnets}
    \end{figure*}

\section{Related Work}
\label{sec:rw}
\subsection{Single Image Super-resolution}
Single image super-resolution (SISR) is a classical problem in the field of image processing. 
To solve this ill-posed inverse problem, a considerable number of regularization-based and example-based methods \cite{sisr2015self,sisr2004ne, sisr2014a+,sisr2008gradient, sisr2010sparse} have been proposed.
Witnessing the great representation ability of deep learning
\cite{krizhevsky2012classification}, Dong \textit{et al.} \cite{sisr2014srcnn,sisr2016srcnn} pioneered deep learning-based methods for SISR, which learn the mapping from  LR to HR images in a data-driven manner.
Later,  deeper network architectures equipped with enhanced feature extraction techniques such as residual and dense connections were widely exploited to improve the SR performance \cite{sisr2016vdsr,sisr2016espcn,sisr2017lapsrn,sisr2018rdn,sisr2017edsr}.
Various loss functions were also proposed to encourage more visually pleasing results, e.g., the perceptual loss \cite{sisr2016perceptual} and the adversarial loss \cite{sisr2017gan}.
More recently, the attention mechanism incorporating non-local operations were introduced to enhance the feature representation and further improve the SR performance \cite{sisr2018rcan,sisr2019san}.
We refer the readers to \cite{sisr2011survey2, sisr2019survey1} for a comprehensive survey on SISR.

\subsection{Reference-based Image Super-resolution}
Reference-based super-resolution (RefSR) utilizes rich and accurate details from a  reference image to assist the SR process.
Benefiting from the extra information provided by the reference image, RefSR can achieve significantly superior performance to  
SISR. 
Zheng \textit{et al.} \cite{refsr2018crossnet, refsr2020crossnet++} proposed to align the feature maps from the reference image to the target LR image via estimating an optical flow. This method requires a high similarity between the reference and LR images, e.g., different views of the same scene in an LF image.
Different from such a global alignment, the idea of local texture matching and transfer was proposed to handle more generic scenarios, where the reference image shares less similar content with the LR image or the correspondences between them have a long distance.
Zhang \textit{et al.} \cite{refsr2019srntt} proposed to search for the matching patches from the reference image in the feature space and then swap the matched features to represent the LR image. Xie \textit{et al.} \cite{refsr2020e2ent} improved this framework by enhancing the feature extractor.
Yang \textit{et al.} \cite{refsr2020ttsr} applied the attention mechanism to transfer and fuse HR features from the reference image into LR features based on their relevance embedding.
Shim \textit{et al.} \cite{refsr2020deformable} utilized stacked deformable convolutional layers equipped with a multi-scale structure and non-local blocks to match similar content between the LR and reference features.
Shao \textit{et al.} \cite{shao2021localtrans} predicted the homography matrix between the cross-resolution image pair in a progressively multi-scale manner, with each scale-level learning a sub-homography from a local attention map by Transformer.
Zhou \textit{et al.} \cite{zhou2021cross} first constructed an SR multi-plane image (MPI) at the LR view and then generated the final SR image by fusing the coarse SR result synthesized from the MPI and the bicubic-upsampled LR image via a CNN.

These RefSR methods can be directly applied to reconstruct an HR LF image from a hybrid input by super-resolving each LR view individually. 
However, it is difficult to preserve the LF structure as the consistency between the reconstructed views is not considered.

\subsection{LF Image Super-resolution}
Different from SISR, LF image super-resolution aims at simultaneously increasing the spatial resolution of all sub-aperture images (SAIs) in an LF image. On top of the target to recover high-frequency details for each SAI, LF super-resolution should also maintain the LF parallax structure. To characterize the relation between SAIs, many methods define a physical model to reconstruct the observed LR SAIs using the desired HR ones. Afterwards, the inverse problem is solved by different priors \cite{lfssr2012gmm,lfssr2014variational,lfssr2014rpca,lfssr2017graph}. These methods always require accurate disparity estimation, which is challenging.

Recent years have witnessed progress on learning-based methods for LF super-resolution. Farrugia \textit{et al.} \cite{lfssr2017subspace} constructed a training set by 3D-stacks of  2-D-patches cropped from different SAIs of paired LF images, and then learned a linear mapping between the subspace of the LR and HR patch stacks.
Yoon \textit{et al.} \cite{lfssr2015yoon} is the first to apply convolutional neural network (CNN) on LF images. However each SAI of an LF image is processed independently in their network, which neglects the angular relationship. Therefore, Yuan \textit{et al.} \cite{lfssr2018combine} proposed to refine the result after separately applying an SISR approach on each SAI.
For the same purpose of keeping the geometric consistency in the reconstructed LF image, Wang \textit{et al.} \cite{lfssr2018recurrent} adopted a recurrent neural network to learn the relations between adjacent SAIs along horizontal and vertical directions. 
To take advantage of the complementary information between SAIs introduced by the LF structure and address the high-dimensionality challenging, Yeung \textit{et al.} \cite{lfssr2018separable} proposed to use 4-D convolution and more efficient spatial-angular separable convolution (SAS-conv) on LF images.
More recently,
Wang \textit{et al.} \cite{lfssr2020interaction} proposed the spatial-angular interaction module to repetitively incorporate spatial and angular information. 
Jin \textit{et al.} \cite{lfssr2020cvpr} proposed an All-to-One module to fuse the combinatorial geometry embedding between the target and auxiliary views in the LF image.

\subsection{LF Image Super-resolution with a Hybrid Input}

LF hybrid imaging system was first proposed by Lu \textit{et al.} \cite{lfhybrid2013microscopy}, in which an HR RGB camera is co-located with a Stack-Hartmann sensor.
Boominathan \textit{et al.} \cite{lfhybrid2014iccp} proposed a patch-based method named \textit{PaSR} to improve the resolution with the hybrid input. Based on \textit{PaSR}, Wang \textit{et al.} \cite{lfhybrid2017ring} improved the performance by iterating between patch-based super-resolution and depth-based synthesis, where the synthesized images were used to update the patch dictionary. The patch-based approaches avoid the need to calibrate and register the DSLR camera and the LF camera. However, the average aggregation causes blurring. Zhao \textit{et al.} \cite{lfhybrid2018tci} proposed a method named \textit{HCSR} to separate the high-frequency details from the HR image and warp them to all SAIs to reconstruct an HR LF image. 
Besides spatial super-resolution, the hybrid LF imaging system was also used to generate LF videos \cite{lfhybrid2017video}.

\section{Proposed Framework}
\label{sec:proposed method}

\textit{Notation}.
Let $\mathcal{L}=\left\{I_\mathbf{u}\in\mathbb{R}^{H\times W}|\mathbf{u}\subset\mathcal{U}\right\}$ denote an LF image with $M\times N$ views of resolution $H\times W$, $\mathcal{U}$ be the set of 2-D angular coordinates of the views, i.e., $\mathcal{U}=\{\mathbf{u}|\mathbf{u}=(u,v),1\leq u\leq M, 1\leq v\leq N\}$, and
$I_{\mathbf{u}}$ denotes the SAI at $\mathbf{u}$.

\subsection{Overview}

As shown in Fig. \ref{fig:flowchat}, a typical hybrid LF imaging system captures an HR central view, denoted by $I^h_{\mathbf{u}_0}\in\mathbb{R}^{\alpha H\times \alpha W}$, surrounded by a set of LR side views, denoted by $\mathcal{S}^l=\mathcal{L}^l\setminus I^l_{\mathbf{u}_0}=\left\{I^l_\mathbf{u}\in\mathbb{R}^{H\times W}|\mathbf{u}\subset\overline{\mathcal{U}}\right\}$, where $\mathbf{u}_0$ denotes the angular coordinate of the central view,
$\overline{\mathcal{U}}= \mathcal{U}\setminus \mathbf{u}_0$,
$\alpha$ is the up-sampling scale factor, and  $\setminus$ means the subtraction of sets.
An HR 4-D LF image to be reconstructed is denoted as
$\tilde{\mathcal{L}}^h=\left\{\tilde{I}^h_\mathbf{u}\in\mathbb{R}^{\alpha H\times\alpha W}|\mathbf{u}\subset\mathcal{U}\right\}$, and the corresponding ground-truth one is denoted as $\mathcal{L}^h=\left\{I^h_\mathbf{u}|\mathbf{u}\subset\mathcal{U}\right\}$.
The problem of reconstructing $\tilde{L}^h$ from the hybrid input can be implicitly formulated as
    \begin{equation}
          \tilde{\mathcal{L}}^h = f\left(I^h_{\mathbf{u}_0},  \mathcal{S}^l\right).
    \end{equation}

To reconstruct $\tilde{\mathcal{L}}^h$, the specific properties of the hybrid input $I^h_{\mathbf{u}_0}$ and $\mathcal{S}^l$ have to be fully explored.
Specifically, $I^h_{\mathbf{u}_0}$ with high spatial resolution captures high-frequency details of the scene, while $\mathcal{S}^l$ with multiple observations from different perspectives 
records geometric information.
Moreover, the image characteristics of the real hybrid data, e.g., the relatively large disparity and color inconsistency across 
views, have to be considered.
Considering the powerful representation ability of deep CNNs, we investigate a deep neural network that can well capture the characteristics of the input to learn such a mapping function $f$.

As shown in Fig. \ref{fig:flowchat}, our framework  consists of two sub-networks, namely SR-Net and Warp-Net. To be specific, by learning deep representations from both $\mathcal{S}^l$ and $I^h_{\mathbf{u}_0}$, the SR-Net aims to super-resolve $\mathcal{S}^l$ via fusing the high-frequency information from $I^h_{\mathbf{u}_0}$, (i.e., to equally increase the spatial resolution of all views contained in $\mathcal{S}^l$), leading to an intermediate HR LF image and its corresponding confidence map, while the Warp-Net 
inversely warps $I^h_{\mathbf{u}_0}$ to side views with the disparity estimated from $\mathcal{S}^l$, generating 
another intermediate HR LF image as well as its confidence map. 
Finally, the two intermediate predictions are adaptively fused based on the learned confidence maps such that only their respective advantages can be leveraged  
into a better output. Note that our framework is trained end-to-end.  In what follows, we will introduce the details of the proposed framework as well as comprehensive analyses.

\subsection{SR-Net}
As depicted in Fig. \ref{fig:subnets}, the SR-Net comprises three modules, i.e., LF feature extraction, HR feature extraction, and hybrid feature fusion. The three modules are connected to promote sufficient exploration of the information contained in the hybrid input.

\subsubsection{LF feature extraction}
Multiple observations from different perspectives  
contained in $\mathcal{S}^l$ provide supplementary information of the scene (i.e., details absent at a certain view may be present in another one), which will  be beneficial to the reconstruction quality.
To capture such information, we stack the LR side views $\{I^l_\mathbf{u}\}$ along the feature channel and utilize sequential convolutional layers with dense connections \cite{huang2017densely,sisr2018rdn} to extract LR features. 
Note that we suppress the feature channels of each group of densely-connected features 
using a bottleneck layer to reduce the number of parameters.
The sub-pixel convolutional layer \cite{sisr2016espcn} is applied to up-sample the extracted features to the desired spatial resolution, i.e.,
    \begin{equation}
            \mathcal{F}^l = \mathsf{UP}\left(f_{sr-l}\left(\mathcal{S}^l\right)\right),    
    \end{equation}
where $\mathcal{F}^l=\{F^l_\mathbf{u}|\mathbf{u}\subset\overline{\mathcal{U}}\}$ is the set of LF features for LR side views, $\mathsf{UP}(\cdot)$ is the up-sampling layer, and $f_{sr-l}(\cdot)$ is the feature extraction layers.

\subsubsection{HR feature extraction}
Considering that $I^h_{\mathbf{u}_0}$ contains rich information and high-frequency details of the scene, we utilize sequential convolutional layers to learn the deep representation of the scene information. To propagate such information to side views, we first explicitly repeat the extracted features, i.e.,
    \begin{equation}
            \mathcal{F}^h = \mathsf{REPEAT}\left(f_{sr-h}\left(I^h_{\mathbf{u}_0}\right)\right),     
    \end{equation}
where $\mathcal{F}^h = \{F^h_\mathbf{u}|\mathbf{u}\subset\overline{\mathcal{U}}\}$ is the extracted HR features, $\mathsf{REPEAT}(\cdot)$ is the repeat operation, and $f_{sr-h}(\cdot)$ is the feature extraction layers.

\subsubsection{Hybrid feature fusion}
For each individual view $I^l_\mathbf{u}$, we combine its LR LF feature and the HR feature via concatenation $\mathsf{CAT}(\cdot)$, then apply convolutional layers with dense connections $f_{sr-f}(\cdot)$ to learn a residual map, denoted as $R^{sr}_\mathbf{u}$:
    \begin{equation}
            R^{sr}_\mathbf{u} = f_{sr-f}\left(\mathsf{CAT}\left(F^l_\mathbf{u}, F^h_\mathbf{u}\right)\right).
    \end{equation}
Finally, we add the residual map to upsampled LR view by the bicubic interpolation $\mathsf{BIC}(\cdot)$ to reconstruct the HR view, i.e.,
    \begin{equation}
            \tilde{I}^{sr}_\mathbf{u} = R^{sr}_\mathbf{u} + \mathsf{BIC}\left(I^l_\mathbf{u}\right),     
    \end{equation}
which constructs the intermediate super-resolved LF by SR-Net, i.e., $\tilde{\mathcal{L}}^{sr}=\{\tilde{I}^{sr}_\mathbf{u}| \mathbf{u}\subset\mathcal{U}\}$.

The SR-Net is trained by minimizing the absolute error between $\tilde{\mathcal{L}}^{sr}$ and the ground-truth HR LF images:
    \begin{equation}
        \begin{aligned}
            \ell^{sr} = \sum_{\mathbf{u}} \sum_{\mathbf{x}} \left|I^h_\mathbf{u}(\mathbf{x}) - \tilde{I}^{sr}_\mathbf{u}(\mathbf{x})\right|.
        \end{aligned}
    \end{equation}

\textit{Remark}.
This module relies on the powerful modeling capacity of the deep CNN to super-resolve $\mathcal{S}^l$ for an intermediate HR LF image. By combining features extracted from $\mathcal{S}^l$ and $I^h_{\mathbf{u}_0}$ for the learning of HR residuals, it is expected that the SR-Net can reconstruct the HR LF image as well as possible. However, its output still suffers from blurry effects caused by the $\ell_1$ loss \cite{video2015beyondMSE,sisr2016perceptual}, although $I^h_{\mathbf{u}_0}$ contains the high-frequency information of the scene.
Additionally, convolutional layers may have difficulties transferring the high-frequency information from $I^h_{\mathbf{u}_0}$ to $\mathcal{S}^l$, because the local operation may be insufficient to cover the large disparity between them.
See the analysis in Sec. \ref{sec:ablation} and Fig. \ref{fig:ablation}. In other words, the high-frequency information embedded in $I^h_{\mathbf{u}_0}$ cannot be very effectively propagated to the output of the SR-Net. To this end, we further develop the following Warp-Net.

\subsection{Warp-Net}
\label{sec:Warp-Net}
As illustrated in Fig. \ref{fig:subnets}, there are two phases involved in this sub-network, i.e.,  disparity estimation and inverse warping. The Warp-Net first learns an HR disparity map for each view by exploring the unique LF structure of $\mathcal{S}^l$ and combining the HR information of $I^h_{\mathbf{u}_0}$, and the resulting HR disparity map is further used to inversely warp $I^h_{\mathbf{u}_0}$, leading to another intermediate HR LF image as well as its confidence map.

\subsubsection{Disparity estimation}
In this phase, we estimate the disparity maps of the LF image by exploring the view relation, i.e., the LF structure embedded in the LR side views.
Specifically,  
under the Lambertian assumption and in the absence of occlusions, such a relation can be expressed as 
    \begin{equation}
            I^l_\mathbf{u}(\mathbf{x}) = I^l_{\mathbf{u}'} \left(\mathbf{x} + d\left(\mathbf{u}'-\mathbf{u}\right)\right), \label{eq:lfstructure}
    \end{equation}
where $d$ is the disparity of point $I^l_\mathbf{u}(\mathbf{x})$.
We use a network with the U-Net structure \cite{ronneberger2015unet}, denoted as $f_{warp-di}(\cdot)$, to exploit the view correlations in $\mathcal{S}^l$, and the output is upsampled to generate the initial disparity map, denoted as $\mathcal{D}^{init}=\left\{D^{init}_\mathbf{u}|\mathbf{u}\subset\overline{\mathcal{U}}\right\} $:
    \begin{equation}
            \mathcal{D}^{init} = \mathsf{UP}\left(f_{warp-di}\left(\mathcal{S}^l\right)\right).
    \end{equation}

$\mathcal{D}^{init}$ roughly describes the scene geometry but lacks high-frequency details to warp the HR central view.
Therefore, we further refine $\mathcal{D}^{init}$ by combining the HR information from $I^h_{\mathbf{u}_0}$.
We utilize sequential convolutional layers to extract features from $\mathcal{D}^{init}$ and $I^h_{\mathbf{u}_0}$, producing $\mathcal{F}^d$ and $\mathcal{F}^{h'}$, respectively.
The extracted geometry and image features are combined via concatenation and then fused using densely-connected convolutional layers denoted as $f_{warp-f}(\cdot)$ to reconstruct residual maps for $\mathcal{D}^{init}$ at individual views, i.e.,
    \begin{equation}
            R^{d}_\mathbf{u} = f_{warp-f}\left(\mathsf{CAT}\left(F^d_\mathbf{u},F^{h'}_\mathbf{u}\right)\right).
    \end{equation}
Finally, we estimate the HR disparity map denoted as $\mathcal{D}^h=\{D^h_\mathbf{u}|\mathbf{u}\subset\overline{\mathcal{U}}\}$ as
    \begin{equation}
            D^{h}_\mathbf{u} = R^d_\mathbf{u} + D^{init}_\mathbf{u}.
    \end{equation}

\subsubsection{Inverse warping}
Based on $\mathcal{D}^h$, another intermediate HR LF image,  denoted as $ \tilde{\mathcal{L}}^{warp}=\left\{\tilde{I}^{warp}_\mathbf{u}|\mathbf{u}\subset\mathcal{U}\right\}$, can be synthesized by inversely warping $I^h_{\mathbf{u}_0}$ to each viewpoint. To make this module end-to-end trainable, we employ the differentiable bicubic interpolation \cite{STN2015} to realize the process of inverse warping:
 
    \begin{equation}
            \tilde{I}^{warp}_\mathbf{u} = \mathsf{WARP}\left(I^h_{\mathbf{u}_0}, D^h_\mathbf{u}, \mathbf{u}-\mathbf{u}_0\right).
    \end{equation}

To train the Warp-Net, we minimize the absolution error between the synthesized HR LF image $\tilde{\mathcal{L}}^{warp}$ and its ground-truth, i.e.,
    \begin{equation}
        \begin{aligned}
            \ell^{warp} = \sum_{\mathbf{u}} \sum_{\mathbf{x}} \left|I^h_\mathbf{u}(\mathbf{x}) - \tilde{I}^{warp}_\mathbf{u}(\mathbf{x})\right|.
        \end{aligned}
    \end{equation}
Moreover, we use an edge-aware smoothness loss \cite{tomasi1998bilateral,jonschkowski2020matters} to regularize the estimated disparity map, i.e.,
    \begin{equation}
        \begin{aligned}
            \ell^{smooth} = \frac{1}{2} \sum_{\mathbf{u}} \sum_{\mathbf{x}} \mathsf{Exp}\left(-\lambda\left|\frac{\partial I^h_\mathbf{u}}{\partial x}(\mathbf{x})\right| \right)\left|\frac{\partial D^h_\mathbf{u}}{\partial x}(\mathbf{x})\right| \\
            + \mathsf{Exp}\left(-\lambda\left|\frac{\partial I^h_\mathbf{u}}{\partial y}(\mathbf{x})\right| \right)\left|\frac{\partial D^h_\mathbf{u}}{\partial y}(\mathbf{x})\right|,
        \end{aligned}
    \end{equation}
where the edge weight $\lambda$ is set to $150$ according to \cite{jonschkowski2020matters}.

\textit{Remark}.
By reusing pixels from $I^h_{\mathbf{u}_0}$, we expect the high-frequency details of the scene that are challenging to predict can be directly transferred from $I^h_{\mathbf{u}_0}$ to each view of $\tilde{\mathcal{L}}^{warp}$. For example, for regions with continuous depths and complicated textures, Warp-Net performs quite well.  See the visual results in Figure \ref{fig:ablation}. However, $\tilde{\mathcal{L}}^{warp}$ inevitably has distortion caused by inaccurate disparity estimations or occlusions. Specifically, it is difficult to obtain accurate disparities without the ground-truth disparities for supervision, especially in challenging regions, such as textureless regions. Such inaccurate disparities will warp pixels of $I^h_{\mathbf{u}_0}$ to wrong positions, resulting in distortion. Second, pixels observed in views of $I^l_\mathbf{u}$ but occluded in $I^h_{\mathbf{u}_0}$ will be occupied by the occluder after warping, causing errors. Interestingly, the SR-Net suffers less from the distortion induced by these two factors. For example, the textureless regions, where the disparities cannot be accurately estimated, correspond to low-frequency contents, which can be relatively easily predicted by the SR-Net. Besides, the powerful regression ability of the SR-Net can predict the occluded pixels to some extent \cite{lfasr2016siggraph}.

\subsection{Confidence-Guided Fusion}
As aforementioned, the SR-Net is capable of predicting the overall content of an HR LF image but fails to recover its delicate textures and sharp edges, while the Warp-Net is able to propagate the high-frequency information to all views but suffers from the distortion caused by occlusions and inaccurate disparity estimation. Fortunately, their advantages are complementary to each other. Therefore, we finally reconstruct an  HR LF image   
by adaptively fusing $\tilde{\mathcal{L}}^{sr}$ and $\tilde{\mathcal{L}}^{warp}$, in which their advantages are leveraged. And such an adaptive fusion process is achieved under the guidance of their own pixel-wise confidence maps.

Both confidence maps are learned from the features extracted by the SR-Net and Warp-Net. 
Specifically, we first use an additional layer parallel to the output layer at the last level to generate the confidence maps denoted as $\mathcal{C}^{sr}=\left\{C_{\mathbf{u}}^{sr}\in\mathbb{R}^{\alpha H\times\alpha W}|\mathbf{u}\subset\overline{\mathcal{U}}\right\}$ and $\mathcal{C}^{warp}  =\left\{C_{\mathbf{u}}^{warp}\in\mathbb{R}^{\alpha H\times\alpha W}|\mathbf{u}\subset\overline{\mathcal{U}}\right\}$ for the SR-Net and Warp-Net, respectively, and then apply the Softmax normalization across $\mathcal{C}^{sr}$ and $\mathcal{C}^{warp}$, generating $\tilde{\mathcal{C}}^{sr} =\left\{\tilde{C}_{\mathbf{u}}^{sr}\in\mathbb{R}^{\alpha H\times\alpha W}|\mathbf{u}\subset\overline{\mathcal{U}}\right\}$ and  $\tilde{\mathcal{C}}^{warp} =\left\{\tilde{C}_{\mathbf{u}}^{warp}\in\mathbb{R}^{\alpha H\times\alpha W}|\mathbf{u}\subset\overline{\mathcal{U}}\right\}$.
The final reconstruction $\tilde{\mathcal{L}}^h$ is produced by the weighted sum of $\tilde{\mathcal{L}}^{sr}$ and $\tilde{\mathcal{L}}^{warp}$: 
    \begin{equation}
        \begin{aligned}
         \tilde{I}^h_\mathbf{u} = \tilde{I}^{sr}_\mathbf{u} \odot  \tilde{C}^{sr}_\mathbf{u} + \tilde{I}^{warp}_\mathbf{u} \odot \tilde{C}^{warp}_\mathbf{u},
        \end{aligned}
    \end{equation} 
where $\odot$ is the element-wise multiplication operator. Such an adaptive fusion process is trained under the supervision of minimizing the $\ell_1$ distance between the final reconstructed HR LF image and  the ground truth one:
    \begin{equation}
        \begin{aligned}
            \ell^{fusion} = \sum_{\mathbf{u}} \sum_{\mathbf{x}} \left|I^h_\mathbf{u}(\mathbf{x}) - \tilde{I}^{h}_\mathbf{u}(\mathbf{x})\right|.
        \end{aligned}
    \end{equation}
Combining all modules, we train the whole network  end-to-end with the following loss function:
    \begin{equation}
        \begin{aligned}
            \ell =   \ell^{fusion} +   \ell^{sr} + \ell^{warp} + \gamma   \ell^{smooth},
        \end{aligned}
    \end{equation}
where the weight factor for smoothness loss $\gamma$ is empirically set to 0.1.

\section{Experiments}
\label{sec:exp}
        \begin{figure}[t]
    \begin{center}
    \includegraphics[width=0.7\linewidth]{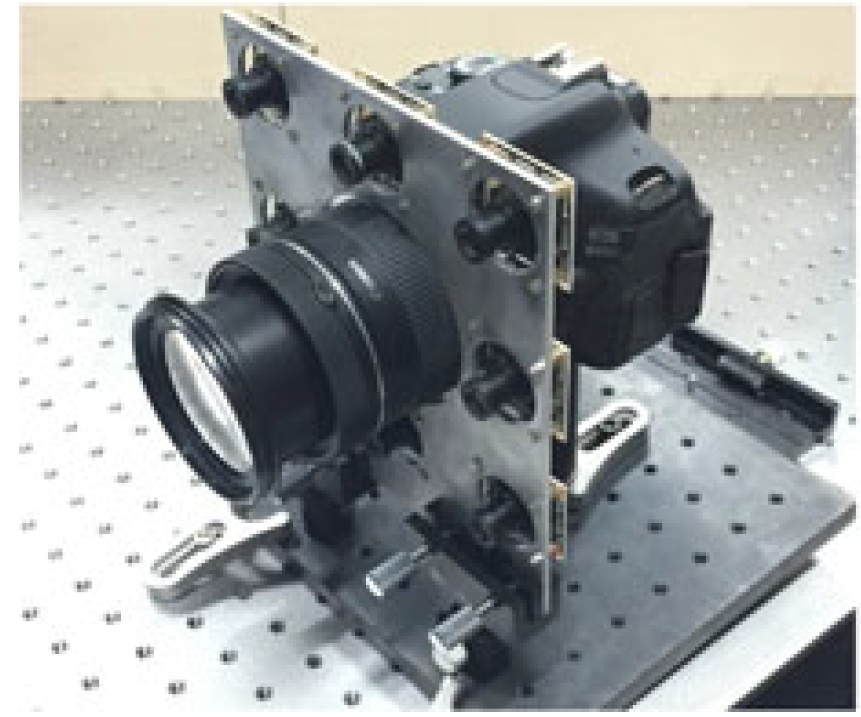}
    \end{center}
    \vspace{-1em}
      \caption{
     Illustration of the prototype of the hybrid LF imaging system built in \cite{lfhybrid2017ring}. This figure is by courtesy of  \cite{lfhybrid2017ring}.
          }
    \label{fig:prototype}
    \end{figure}
    \begin{figure*}[t]
    \begin{center}
    \includegraphics[width=0.9\linewidth]{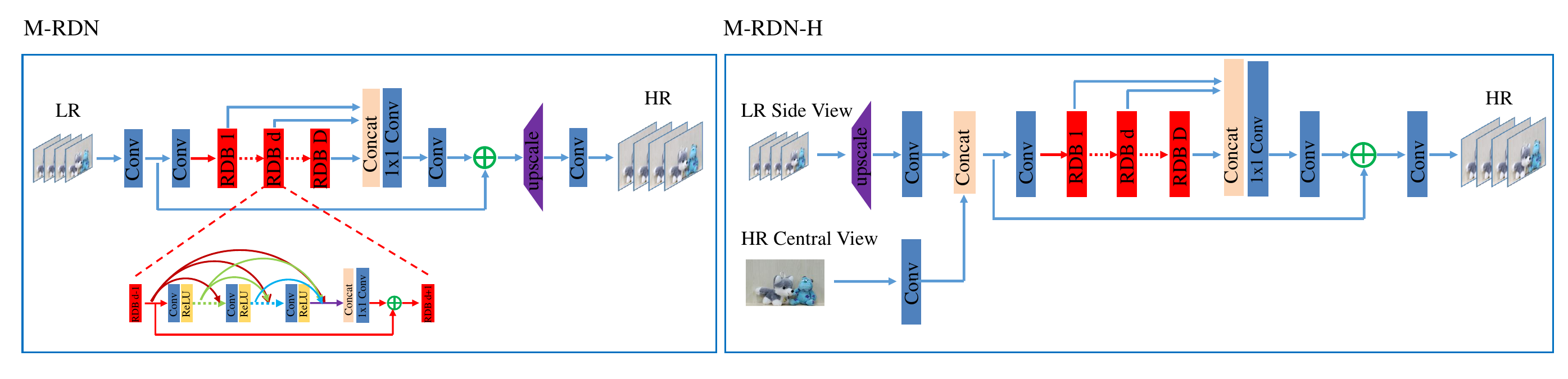}
    \end{center}
    \vspace{-1em}
    \caption{
    Illustration of the network architectures of two newly developed baseline methods named M-RDN and M-RDN-H that are built upon RDN \cite{sisr2018rdn}.
    }
    \label{fig:mrdnh}
    \end{figure*}

    \begin{figure*}[thp]
    \begin{center}
    \includegraphics[width=1.0\linewidth]{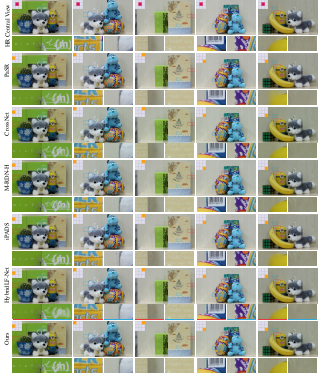}
    \end{center}
    \vspace{-1em}
    \caption{
    Visual comparisons of different methods on real hybrid data. For each algorithm, we provide the zoom-in images of the red and blue blocks. The colored grid on the top-left corner of each image indicates its angular position.
    }
    \label{fig:real}
    \end{figure*}

    \begin{figure*}[t]
    \begin{center}
    \includegraphics[width=0.9\linewidth]{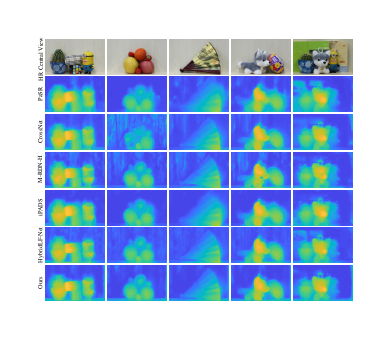}
    \end{center}
    \vspace{-0.3cm}
    \caption{
    Visual comparisons of the estimated depth maps from the reconstructed HR LF images by different methods on real hybrid data.
    }
    \label{fig:real_disp}
    \end{figure*}

    \begin{table}[t]
    \centering
    \caption{Comparisons of the average running time (in seconds) and the number of parameters (\#Params) of different methods for reconstructing an HR LF image from real hybrid data. 
    Note that learning-based methods, i.e., CrossNet, M-RDN-H, HybridLF-Net, and Ours, require running 3 times to reconstruct individual channels of the image in YCbCr color space.
    }
    \label{tab:time}
    \resizebox{ \linewidth}{!}{
    \begin{tabular}{c|c c c c c c}
    \toprule
              & PaSR &  CrossNet & M-RDN-H & iPADS & HybridLF-Net   & Ours\\
    \midrule
    Time      & 722.23s    & 12.80s    & 6.45s   & 7385.37s & 15.74s& 14.85s \\
    \#Params. & -    & 35.16M    & 22.06M  & -   & 2.32M    & 10.21M\\
    \bottomrule
    \end{tabular}
    }
    \end{table}

    \begin{figure*}[t]
    \begin{center}
    \includegraphics[width=0.9\linewidth]{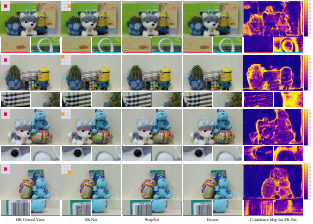}
    \end{center}
    \vspace{-1.2em}
      \caption{
      Visual comparisons of intermediate predictions by the SR-Net and Warp-Net. Note that as the sum of the confidence maps of the SR-Net and Warp-Net is equal to 1, we only visualized the confidence map of the SR-Net. 
      The zoomed-in \textcolor{red}{Red} frames highlight the advantages of Warp-Net, while the zoomed-in \textcolor{blue}{Blue} frames highlight the advantages of SR-Net.}
    \label{fig:ablation}
    \end{figure*}

    \begin{figure*}[t]
    \begin{center}
    \includegraphics[width=0.9\linewidth]{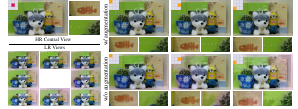}
    \end{center}
    \vspace{-1.2em}
      \caption{Visual comparisons of the reconstruction results by our method trained w/ and w/o color augmentation. We also provided the HR central view and LR views for reference. 
      See the associated \textit{video demo} for more results.
      }
    \label{fig:ablation_coloraug}
    \end{figure*}

\subsection{Implementation Details and Data Augmentation}
In our network, we set  the kernel size of all convolutional layers to $3\times3$ except that of the bottleneck layers, whose kernel size is $1\times1$, and applied zero-padding to keep the spatial resolution unchanged. 
During training, we randomly cropped images to patches of spatial resolution $128\times128$. We set the batch size to 1 and used Adam optimizer \cite{kingma2014adam} with $\beta_1=0.9$ and $\beta_2=0.999$. The learning rate was initialized as $1e^{-4}$ and decreased by a factor of 0.5 every 250 epochs.

Due to the limited number of images in current high-quality LF datasets, it is necessary to apply data augmentation to increase the diversity of the training samples.
However,
commonly used methods for data augmentation, including image rotation and flip, do not work for LF data.
Specifically,
if we apply these transformations on each SAI separately, the LF structure in Eq. \ref{eq:lfstructure} would be destroyed. For example, 
applying the flip operation along the $y$ dimension, we have 
    \begin{equation}
        \begin{aligned}
            &I_{u,v}(x, W-y) \\
            & =  I_{u+\Delta u,v+\Delta v} (x+d\Delta u, W-( y + d\Delta v))\\
            & =  I_{u+\Delta u,v+\Delta v} (x+d \Delta u,(W- y)- d\Delta v),
        \end{aligned}
    \end{equation}
where it can be seen that the relation between the flipped view $I_\mathbf{u}$ and $I_{\mathbf{u}+\Delta\mathbf{u}}$  disobeys  
Eq. \ref{eq:lfstructure}. 
Therefore, we propose a new data augmentation method tailored for LF data, i.e., applying the image geometric augmentation methods on angular and spatial dimensions simultaneously. With our new strategy, taking the flip augmentation along the $y$ dimension as an example again, we have
    \begin{equation}
        \begin{aligned}
            & I_{u,N-v}(x,W-y)  \\
            & = I_{u+\Delta u, N-(v+\Delta v)} (x+d\Delta u, W-(y+d\Delta v))  \\
            & = I_{u+\Delta u, (N-v)-\Delta v} (x+d\Delta u, (W-y)-d\Delta v),
        \end{aligned}
    \end{equation}
where the LF structure described in Eq. \ref{eq:lfstructure} still holds  in the flipped LF image.

Moreover, in real hybrid data, different views usually have obvious brightness and color inconsistency due to the change of illumination, camera lens, and viewpoints.
To increase the robustness of the model to color inconsistency across  views, we augmented the training samples by randomly and independently changing the brightness, contrast, saturation, and hue of each side view of the input, while keeping the color of the supervision data unchanged.
We will validate the effectiveness of the color augmentation in Sec. \ref{sec:ablation}.

\subsection{Evaluation on Real Hybrid Data}
\label{sec:real}
\subsubsection{Dataset and Training Strategy}
To evaluate the proposed framework, we adopted the real hybrid data captured by the hybrid LF imaging prototype built in \cite{lfhybrid2017ring}, as shown in Fig. \ref{fig:prototype}, which attaches eight low-cost LR side cameras around a central high-quality HR DSLR camera.
Each scene image captured by this prototype consists of eight low-quality side views of spatial resolution around $900\times 1482$, and a high-quality central view of spatial resolution around $1729\times2846$. 
The intrinsic parameters of the cameras and the extrinsic parameters with reference to the central camera were calibrated using a standard camera calibration toolbox and bundle adjustment software, respectively.  The side-view images were rectified to share the same rotation with the central image by projection and homography warping.  Finally, the  $3\times3$ views are nearly regularly placed on a 2-D plane with the same rotation.
We also refer readers to  \cite{lfhybrid2017ring} for more details about the settings of the prototype.

To learn a model suitable for such real data, we particularly designed the training strategy.
As the ground-truth HR LF images are not available for supervision in the real hybrid dataset, we simulated hybrid data from publicly available LF datasets for training, i.e., 
we spatially down-sampled 
off-center views of  LF images  
from the Inria Dense dataset \cite{lfdataset2019inria} and the HCI benchmark \cite{lfdataset2016hci}, 
which contain synthetic LF images of spatial resolution $512\times 512$, angular resolution of $9\times 9$, and disparity in the range of $[-4, 4]$.
Considering the angular resolution of the real data for testing and the observation that real hybrid data usually have relatively large disparities,  
we also uniformly sampled 
$3\times 3$ SAIs from $9\times 9$ SAIs of LF images, leading 44 simulated hybrid inputs with a disparity range of $[-16, 16]$ for training.

We converted the RGB images to YUV color space and only used the Y component for training.
During testing, to address the problem of color inconsistency across views, we first converted the input hybrid data to YUV color space, and then 
applied the trained model to reconstruct the luminance and two chrominance channels separately.

\subsubsection{Comparison with state-of-the-art methods}
\label{subsubsec:real_sota}
To demonstrate the advantages of the proposed method, we compared it  
with state-of-the-art methods, 
including
two traditional step-wise  methods for LF SR from hybrid inputs, i.e., PaSR \cite{lfhybrid2014iccp} and iPADS \cite{lfhybrid2017ring}, 
a deep learning-based method for LF SR from hybrid inputs, i.e., HybridLF-Net \cite{lfhybrid2020fusion},
and a deep learning-based RefSR method, i.e., CrossNet \cite{refsr2018crossnet}.
Additionally, based on RDN \cite{sisr2018rdn}, a state-of-the-art SISR method, we developed a strong baseline network, namely M-RDN-H, to handle a hybrid LF input.  
Specifically, as shown in Fig. \ref{fig:mrdnh}, M-RDN-H first extracts features from the stacked LR side views and HR central view separately, and then concatenates these feature maps together. The combined features are fed into the network with a similar structure to RDN to learn the mapping from the LR to HR space. 
The LR side views are up-sampled using bicubic interpolation to share the same spatial resolution of the HR view before being fed into the network.
Note that we re-trained all the learning-based methods under comparison using the same training dataset and strategy as \textit{Ours} for \textit{fair} comparisons.
The YUV channels are also processed in the same way for learning-based methods, while 
conventional methods process the images in RGB space directly.

\textbf{Comparison of visual results.}
Fig. \ref{fig:real} provides visual comparisons of the reconstructed LFs by different methods,  
where it can be observed that:
\begin{itemize}  
\item 
all of PaSR \cite{lfhybrid2014iccp}, CrossNet \cite{refsr2018crossnet}, M-RDN-H, and  HybridLF-Net \cite{lfhybrid2020fusion} suffer from serious blurry effects, such as the letters on the notebook cover, the barcode, and the wood texture on the wall. To be specific, PaSR \cite{lfhybrid2014iccp} searches for 9 nearest neighbors in the feature space for each LR patch, and then reconstructs this patch by weighted averaging the corresponding HR patches. Such an average operation causes the loss of the high-frequency details in the HR patches. 
CrossNet \cite{refsr2018crossnet} estimates a flow between the LR and HR input views, which is further used to align the two views  
in  feature space for reconstruction. However, as the flow is predicted between the cross-domain images, i.e., the LR and HR views, without proper guidance, i.e., the prediction process is only driven by the final reconstruction loss,  it is hard to accurately align the HR features to the target view when the disparity increases, resulting in insufficient propagation of the high-frequency details. 
M-RDN-H is a pure regression-based method, whose limited performance could be caused by the relatively large disparity between input views as the local convolutions have difficulties exploring the long-distance correlations.
Besides, as HybridLF-Net \cite{lfhybrid2020fusion} was built  
on simulated hybrid data, it fails to handle the challenges posed by the color inconsistency and large disparity of real hybrid data. Thus, the high-frequency details in the HR view are not effectively propagated to side views.
\item  
iPADS \cite{lfhybrid2017ring}  suffers from distortions around depth discontinuous boundaries.
As iPADS renders HR side views by warping the HR central view based on an estimated depth map, high-frequency details can be preserved relatively well on plain areas. However, this method inevitably causes distortions due to depth inaccuracy and occlusions, as we analyzed in Sec. \ref{sec:Warp-Net}; and
\item our approach produces satisfactory results on both textured areas and occlusion boundaries.
Owing to the confidence-guided fusion framework, the results of our approach keep the high-frequency details explicitly propagated from the HR central view and the geometric structure around occlusion boundaries simultaneously.
\end{itemize}
We refer the readers to the associated \textit{video demo} \footnote {\url{https://github.com/jingjin25/LFhybridSR-Fusion}} for more results.

\textbf{Comparison of the LF parallax structure}.
The most valuable information of LF data is the LF parallax structure as described in Eq. \ref{eq:lfstructure}, which implicitly represents the geometry of the scene/object.
To evaluate the ability of different methods in preserving the LF parallax structure, we visually compared the depth/disparity maps estimated from the reconstructed HR LF images by different methods using an identical LF depth estimation algorithm \cite{lfdepth2018jake}.
Fig. \ref{fig:real_disp} shows the results,  
where it can be observed that our approach can produce much better disparity maps.
Specifically, the disparity maps from PaSR \cite{lfhybrid2014iccp}, and iPADS \cite{lfhybrid2017ring} present obviously blurry around object edges.
The reason is that patch matching and depth-based warping generally cause blurry or distortion around depth discontinuous boundaries, leading to view inconsistency in these areas.
The disparity maps from CrossNet \cite{refsr2018crossnet}, M-RDN-H, and HybridLF-Net \cite{lfhybrid2020fusion} keep sharp edges of the objects, but show obvious errors on areas with weak textures and large disparities, especially the background. 
In contrast, the disparity maps from our approach keep sharper edges and describe more accurate geometry for both foreground objects and backgrounds, demonstrating the stronger ability of our method to preserve the LF parallax structure than other methods.

\subsubsection{Efficiency}
We also compared the computational complexities of different methods by measuring the running time (in seconds) of the testing phase and the number of parameters of deep learning-based methods.  
All methods were tested on a desktop with Intel Xeon Silver  4215R CPU@3.20GHz, 128 GB RAM and NVIDIA Quadro RTX 8000.
As listed in Table \ref{tab:time}, it can be observed that learning-based methods, i.e., CrossNet, M-RDN-H, HybridLF-Net, and Ours, are much faster than conventional methods, i.e., PaSR and iPADS.
Although our approach takes a slightly longer time than CrossNet and M-RDN-H, its model size is much smaller than theirs.
Taking the trade-off between computational complexity and reconstruction quality, we believe our method is competing.

    \begin{table*}[t]
    \centering
    \caption{
    Quantitative comparisons of the proposed approach with state-of-the-art ones on simulated hybrid data.
    PSNR/SSIM/LPIPS over total 19 test LF images on $4\times$ and $8\times$ reconstruction are provided.
    The best and second best results are colored in \textcolor{red}{red} and \textcolor{blue}{blue}, respectively.
    }\vspace{-0.3cm}
    \label{tab:quan_simulate}
    \setlength\tabcolsep{3pt}
    \resizebox{\textwidth}{!}{
    \begin{tabular}{c|c|c c c c c c c|c}
    \toprule[1pt]
Scale & LF             & Bicubic     & SAS-conv \cite{lfssr2018separable}   & M-RDN       & PaSR \cite{lfhybrid2014iccp}        & CrossNet \cite{refsr2018crossnet}    & M-RDN-H     &  HybridLF-Net \cite{lfhybrid2020fusion}   &  Ours\\
\midrule[0.5pt]
\multirow{19}{*}{$4\times$}    
      & Bedroom             & 30.95/0.899/0.493 & 33.87/0.947/0.258 & 33.61/0.943/0.229 & 34.51/0.880/0.229 & 37.80/0.977/0.050 & 39.69/\textcolor{blue}{0.984}/\textcolor{blue}{0.032} & \textcolor{red}{39.95/0.985/0.030} & \textcolor{blue}{39.88}/\textcolor{red}{0.985/0.030} \\
      & Boardgames          & 27.80/0.879/0.350 & 34.01/0.965/0.068 & 33.10/0.956/0.083 & 34.62/0.959/0.093 & 39.35/0.990/\textcolor{blue}{0.022} & \textcolor{blue}{43.47/0.996}/\textcolor{red}{0.014} & 43.40/\textcolor{blue}{0.996}/\textcolor{red}{0.014} & \textcolor{red}{44.38/0.997}/\textcolor{red}{0.014} \\
      & Sideboard           & 24.03/0.743/0.454 & 27.95/0.896/0.189 & 27.52/0.881/0.177 & 26.74/0.799/0.225 & 29.50/0.930/0.074 & \textcolor{red}{34.18}/0.973/\textcolor{red}{0.035} & 33.78/\textcolor{blue}{0.976}/\textcolor{blue}{0.036} & \textcolor{blue}{33.92}/\textcolor{red}{0.977}/\textcolor{red}{0.035} \\
      & Town                & 28.61/0.871/0.417 & 32.13/0.934/0.200 & 31.73/0.928/0.179 & 31.45/0.870/0.209 & 36.58/0.977/0.035 & \textcolor{blue}{40.58}/0.989/\textcolor{red}{0.016} & 40.49/\textcolor{blue}{0.991}/\textcolor{blue}{0.017} & \textcolor{red}{40.92/0.992}/\textcolor{red}{0.016} \\
      & Antiques            & 36.65/0.961/0.312 & 40.13/0.983/0.115 & 40.08/0.983/0.102 & 39.70/0.962/0.122 & 44.10/\textcolor{blue}{0.993}/0.035 & 47.00/\textcolor{red}{0.997}/\textcolor{blue}{0.017} & \textcolor{blue}{47.50}/\textcolor{red}{0.997}/0.018 & \textcolor{red}{48.07/0.997}/\textcolor{red}{0.015} \\
      & Camera\_brush       & 28.73/0.907/0.374 & 33.44/0.961/0.152 & 32.77/0.956/0.156 & 34.51/0.924/0.144 & 37.13/0.978/0.045 & 38.36/\textcolor{red}{0.984}/\textcolor{red}{0.024} & \textcolor{blue}{39.03/0.983}/\textcolor{blue}{0.025} & \textcolor{red}{39.07}/0.982/\textcolor{red}{0.024} \\
      & Chess               & 26.11/0.905/0.471 & 31.30/0.947/0.309 & 29.83/0.938/0.346 & 32.21/0.878/0.285 & 34.42/0.966/0.128 & 37.39/\textcolor{blue}{0.980}/\textcolor{red}{0.036} & \textcolor{blue}{37.68}/\textcolor{red}{0.983}/0.050 & \textcolor{red}{38.83/0.983}/\textcolor{blue}{0.043} \\
      & Coffee\_time        & 21.20/0.670/0.572 & 24.72/0.850/0.257 & 24.33/0.831/0.250 & 26.74/0.882/0.256 & 31.15/0.972/0.057 & \textcolor{blue}{37.46}/0.991/\textcolor{blue}{0.034} & 36.93/\textcolor{blue}{0.992}/0.035 & \textcolor{red}{38.13/0.994}/\textcolor{red}{0.033} \\
      & Flowers\_clock      & 30.98/0.938/0.260 & 36.46/0.980/0.065 & 35.23/0.974/0.080 & 34.95/0.955/0.080 & 39.27/0.989/0.032 & 41.59/\textcolor{blue}{0.993}/0.021 & \textcolor{blue}{41.97}/\textcolor{red}{0.994}/\textcolor{blue}{0.021} & \textcolor{red}{42.37/0.994}/\textcolor{red}{0.020} \\
      & Lonely\_man         & 30.33/0.927/0.246 & 33.29/0.960/0.111 & 32.95/0.956/0.116 & 33.88/0.936/0.106 & 36.28/0.982/0.034 & 33.78/0.974/0.023 & \textcolor{blue}{37.75/0.987}/\textcolor{blue}{0.023} & \textcolor{red}{38.83/0.989}/\textcolor{red}{0.021} \\
      & Microphone\_rooster & 23.58/0.831/0.375 & 27.25/0.916/0.103 & 26.47/0.905/0.143 & 29.36/0.921/0.102 & 31.20/0.972/0.040 & \textcolor{blue}{35.80/0.986}/\textcolor{blue}{0.014} & 35.36/0.985/0.017 & \textcolor{red}{36.82/0.987}/\textcolor{red}{0.013} \\
      & Pinenuts\_blue      & 29.97/0.859/0.465 & 32.91/0.913/0.264 & 32.11/0.902/0.234 & 34.13/0.875/0.209 & 36.82/0.967/0.056 & \textcolor{blue}{39.13/0.980}/\textcolor{red}{0.026} & 38.05/0.975/0.045 & \textcolor{red}{39.53/0.982}/\textcolor{blue}{0.030} \\
      & Rooster\_clock      & 28.75/0.875/0.499 & 34.02/0.955/0.215 & 33.29/0.948/0.187 & 33.79/0.894/0.235 & 40.50/0.989/0.044 & \textcolor{blue}{44.91/0.995}/0.019 & 44.82/\textcolor{red}{0.996}/\textcolor{blue}{0.019} & \textcolor{red}{45.49/0.996}/\textcolor{red}{0.018} \\
      & Roses\_bed          & 30.49/0.925/0.332 & 33.12/0.963/0.116 & 33.13/0.961/0.118 & 37.20/0.957/0.103 & 34.73/0.977/0.041 & 35.39/\textcolor{red}{0.983}/\textcolor{red}{0.027} & \textcolor{red}{36.22/0.983}/\textcolor{red}{0.027} & \textcolor{blue}{35.52/0.981}/\textcolor{blue}{0.028} \\
      & Roses\_table        & 30.08/0.906/0.365 & 33.46/0.955/0.111 & 32.78/0.948/0.149 & 32.99/0.940/0.120 & 35.37/0.970/0.049 & 35.89/0.973/\textcolor{red}{0.031} & \textcolor{red}{36.87/0.977}/0.045 & \textcolor{blue}{36.54/0.975}/\textcolor{blue}{0.038} \\
      & Toy\_friends        & 30.30/0.882/0.498 & 32.01/0.911/0.327 & 31.67/0.907/0.299 & 32.07/0.815/0.297 & 35.54/0.963/0.059 & 37.43/\textcolor{blue}{0.973}/0.032 & \textcolor{blue}{37.69}/\textcolor{red}{0.977}/\textcolor{red}{0.030} & \textcolor{red}{37.91/0.977}/\textcolor{blue}{0.031} \\
      & Toys                & 28.20/0.886/0.509 & 32.39/0.940/0.331 & 31.98/0.937/0.321 & 32.35/0.836/0.279 & 35.91/0.967/0.052 & 39.14/\textcolor{blue}{0.982}/0.021 & \textcolor{blue}{39.18/0.982}/\textcolor{blue}{0.020} & \textcolor{red}{39.43/0.983}/\textcolor{red}{0.019} \\
      & Two\_vases          & 30.97/0.919/0.307 & 34.85/0.961/0.091 & 34.34/0.957/0.082 & 36.56/0.954/0.089 & 40.24/0.989/\textcolor{blue}{0.027} & \textcolor{red}{42.39/0.992}/\textcolor{red}{0.017} & \textcolor{blue}{40.90/0.989}/\textcolor{red}{0.017} & 40.89/\textcolor{blue}{0.989}/\textcolor{red}{0.017} \\
      & White\_roses        & 30.63/0.919/0.382 & 35.35/0.965/0.092 & 34.65/0.961/0.111 & 33.17/0.942/0.126 & 39.20/0.987/0.038 & 41.23/\textcolor{blue}{0.992}/\textcolor{red}{0.019} & \textcolor{blue}{41.64/0.992}/\textcolor{blue}{0.023} & \textcolor{red}{42.04/0.993}/\textcolor{red}{0.019} \\
      \midrule[0.5pt]
\multicolumn{2}{c|}{Avg.}   & 28.86/0.879/0.404 & 32.77/0.942/0.178 & 32.18/0.936/0.177 & 33.20/0.905/0.174 & 36.58/0.975/0.048 & 39.20/0.985/\textcolor{red}{0.024} & \textcolor{blue}{39.43/0.986}/\textcolor{blue}{0.027} & \textcolor{red}{39.91/0.987}/\textcolor{red}{0.024} \\
\midrule[0.5pt]
\multirow{19}{*}{$8\times$}    
      & Bedroom             & 28.38/0.845/0.713 & 30.45/0.905/0.490 & 30.29/0.898/0.454 & 33.18/0.852/0.265 & 35.81/0.966/0.099 & 36.58/\textcolor{blue}{0.977}/\textcolor{blue}{0.065} & \textcolor{blue}{37.51}/0.975/0.069 & \textcolor{red}{37.80/0.978}/\textcolor{red}{0.063} \\
      & Boardgames          & 24.47/0.781/0.639 & 28.32/0.890/0.311 & 27.76/0.877/0.354 & 31.79/0.923/0.131 & 35.32/0.980/0.070 & 34.16/0.968/\textcolor{blue}{0.056} & \textcolor{blue}{38.38/0.990}/0.058 & \textcolor{red}{40.67/0.994}/\textcolor{red}{0.054} \\
      & Sideboard           & 21.20/0.590/0.697 & 23.02/0.747/0.425 & 22.97/0.736/0.447 & 24.21/0.650/0.282 & 25.91/0.849/0.162 & 28.55/0.925/\textcolor{red}{0.124} & \textcolor{blue}{29.30/0.932}/0.133 & \textcolor{red}{29.90/0.942}/\textcolor{blue}{0.126} \\
      & Town                & 25.67/0.794/0.647 & 28.62/0.886/0.372 & 28.26/0.876/0.420 & 29.40/0.827/0.227 & 33.64/0.958/0.099 & \textcolor{blue}{36.78/0.978}/\textcolor{red}{0.063} & 35.71/0.974/0.067 & \textcolor{red}{36.98/0.981}/\textcolor{blue}{0.066} \\
      & Antiques            & 33.38/0.930/0.547 & 35.50/0.953/0.336 & 35.63/0.954/0.306 & 36.79/0.934/0.185 & 40.41/0.984/0.093 & \textcolor{blue}{44.17}/\textcolor{red}{0.994}/\textcolor{red}{0.056} & 42.77/\textcolor{blue}{0.991}/\textcolor{blue}{0.071} & \textcolor{red}{44.25/0.994}/\textcolor{red}{0.056} \\
      & Camera\_brush       & 25.28/0.843/0.556 & 28.63/0.917/0.327 & 28.54/0.911/0.356 & 32.77/0.903/0.181 & 35.28/0.971/0.084 & 36.25/0.978/\textcolor{blue}{0.053} & \textcolor{blue}{37.35/0.979}/0.057 & \textcolor{red}{38.35/0.983}/\textcolor{red}{0.052} \\
      & Chess               & 22.99/0.846/0.605 & 26.55/0.918/0.373 & 25.81/0.907/0.437 & 29.99/0.868/0.325 & 30.39/0.951/0.241 & \textcolor{blue}{33.79}/0.959/\textcolor{red}{0.056} & 33.21/\textcolor{blue}{0.960}/\textcolor{blue}{0.135} & \textcolor{red}{35.76/0.974}/\textcolor{red}{0.056} \\
      & Coffee\_time        & 18.85/0.466/0.807 & 20.45/0.611/0.564 & 20.18/0.582/0.632 & 24.98/0.837/0.295 & 28.02/0.941/0.156 & 31.41/0.975/\textcolor{blue}{0.126} & \textcolor{blue}{32.82/0.979}/0.131 & \textcolor{red}{34.53/0.987}/\textcolor{red}{0.123} \\
      & Flowers\_clock      & 27.22/0.882/0.412 & 31.54/0.953/0.174 & 30.88/0.943/0.226 & 32.39/0.934/0.111 & 36.20/0.980/0.059 & \textcolor{blue}{38.60/0.989}/\textcolor{blue}{0.041} & 37.92/0.987/0.046 & \textcolor{red}{39.81/0.990}/\textcolor{red}{0.040} \\
      & Lonely\_man         & 27.44/0.880/0.391 & 30.16/0.931/0.221 & 29.82/0.923/0.272 & 31.33/0.908/0.141 & 34.13/0.972/0.096 & 34.50/0.977/\textcolor{blue}{0.074} & \textcolor{blue}{38.20/0.986}/0.079 & \textcolor{red}{39.07/0.989}/\textcolor{red}{0.071} \\
      & Microphone\_rooster & 19.17/0.655/0.628 & 23.23/0.825/0.268 & 22.51/0.801/0.364 & 25.24/0.845/0.171 & 25.40/0.934/0.085 & 29.17/0.949/\textcolor{red}{0.048} & \textcolor{blue}{29.72/0.967}/\textcolor{blue}{0.058} & \textcolor{red}{30.72/0.974}/\textcolor{red}{0.048} \\
      & Pinenuts\_blue      & 27.30/0.782/0.675 & 29.80/0.848/0.466 & 29.20/0.834/0.500 & 31.96/0.831/0.247 & 33.70/0.938/0.132 & 35.79/0.956/\textcolor{blue}{0.090} & \textcolor{blue}{36.01/0.963}/0.098 & \textcolor{red}{36.57/0.969}/\textcolor{red}{0.086} \\
      & Rooster\_clock      & 25.33/0.784/0.688 & 28.06/0.883/0.456 & 27.96/0.873/0.498 & 32.37/0.869/0.263 & 37.77/0.981/0.105 & \textcolor{blue}{41.03/0.991}/\textcolor{red}{0.071} & 40.82/0.990/\textcolor{blue}{0.077} & \textcolor{red}{42.55/0.993}/\textcolor{red}{0.071} \\
      & Roses\_bed          & 28.56/0.886/0.518 & 30.08/0.927/0.260 & 30.21/0.923/0.296 & 33.28/0.929/0.142 & 34.25/0.974/0.066 & 31.08/0.960/0.045 & \textcolor{blue}{38.23/0.987}/\textcolor{red}{0.039} & \textcolor{red}{39.59/0.990}/\textcolor{blue}{0.040} \\
      & Roses\_table        & 27.04/0.841/0.620 & 30.18/0.924/0.264 & 29.31/0.913/0.345 & 29.65/0.878/0.203 & 33.59/0.963/0.091 & 32.51/0.957/\textcolor{blue}{0.062} & \textcolor{blue}{35.59/0.976}/0.078 & \textcolor{red}{35.96/0.979}/\textcolor{red}{0.060} \\
      & Toy\_friends        & 28.50/0.848/0.643 & 30.17/0.882/0.495 & 29.84/0.877/0.479 & 31.17/0.794/0.328 & 33.57/0.943/0.115 & 35.11/0.960/\textcolor{blue}{0.056} & \textcolor{blue}{35.44/0.964}/0.060 & \textcolor{red}{36.44/0.971}/\textcolor{red}{0.050} \\
      & Toys                & 25.48/0.833/0.651 & 28.00/0.898/0.462 & 27.85/0.891/0.500 & 30.95/0.812/0.321 & 33.17/0.942/0.137 & 33.99/0.945/\textcolor{blue}{0.054} & \textcolor{blue}{35.46/0.958}/0.074 & \textcolor{red}{36.22/0.969}/\textcolor{red}{0.048} \\
      & Two\_vases          & 27.59/0.850/0.516 & 29.51/0.898/0.311 & 29.42/0.892/0.330 & 34.39/0.929/0.121 & 37.17/0.979/0.065 & 38.84/0.984/\textcolor{blue}{0.046} & \textcolor{blue}{38.93/0.985}/0.048 & \textcolor{red}{40.05/0.988}/\textcolor{red}{0.045} \\
      & White\_roses        & 26.91/0.841/0.630 & 29.78/0.904/0.309 & 29.47/0.897/0.365 & 30.65/0.900/0.183 & 33.35/0.955/0.091 & \textcolor{red}{35.27/0.966}/\textcolor{red}{0.056} & \textcolor{blue}{34.80/0.965}/0.073  & 34.26/0.957/\textcolor{blue}{0.061} \\
      \midrule[0.5pt]
      \multicolumn{2}{c|}{Avg.} & 25.83/0.799/0.610 & 28.53/0.879/0.362 & 28.20/0.869/0.399 & 30.87/0.864/0.217 & 33.53/0.955/0.108 & 35.13/0.968/\textcolor{blue}{0.066} & \textcolor{blue}{36.22/0.974}/0.076 & \textcolor{red}{37.34/0.979}/\textcolor{red}{0.064}  \\
    \bottomrule[1pt]
    \end{tabular}
    }
    \end{table*}

\subsubsection{Ablation study}
\label{sec:ablation}
Here, we provided ablation studies to validate the effectiveness of the framework and the training strategy.

\textbf{Effectiveness of the fusion manner}. 
To investigate the difference between the SR-Net and Warp-Net and their contributions to the final output, and consequently validate the effectiveness of the fusion component,
we visually compared the intermediate predictions by SR-Net and Warp-Net, the corresponding confidence maps, and the final output.
As shown in Fig. \ref{fig:ablation}, it can be seen that 
for plain areas ( highlighted in \textit{red} frames), the SR-Net produces seriously blurry results and fails to recover the textured details, while the Warp-Net can accurately propagate the high-frequency textures from the HR input view. The confidence maps also show that the Warp-Net has higher weights for the final reconstruction in these areas.
For areas with discontinuous depth (highlighted in \textit{blue} frames), 
the predictions of the Warp-Net have distortions while those of the SR-Net maintain the content and provide more contributions to the final outputs. 
Therefore, we can conclude that the SR-Net and Warp-Net present advantages separately in different areas, and the fusion component is indeed able to leverage the advantages of these two modules to produce better final results.

\textbf{Effectiveness of the color augmentation.} We compared the reconstruction results of our method trained with (w/) and without (w/o) the color augmentation strategy.
As shown in Fig. \ref{fig:ablation_coloraug}, it can be seen that our method  trained w/o color augmentation produces blurry results with color inconsistency in both spatial and angular domains. 
More specifically,
the colors of SR-Net's results are mainly influenced by the variant color of the individual side view, while the colors of Warp-Net's results are the same as that of the HR central view. Consequently, the  confidence-based fusion results show inconsistent colors inside each view and cross different views.
In contrast, the results by our method trained w/ color augmentation preserve intra-view high-frequency details and inter-view color consistency, demonstrating the effectiveness of the color augmentation strategy.

\subsection{Evaluation on Simulated Hybrid Data}
\label{sec:simulate}
To have a quantitative understanding of the advantages of our method, here we also conducted evaluations on simulated hybrid data, which can provide ground-truth HR LF images, although there is a significant gap between real and simulated hybrid data, such as the large disparity and color inconsistency cross views.

\subsubsection{Datasets and training details} 
We generated simulated hybrid data 
by down-sampling off-center views of an LF image and only retaining  the resolution of the central view.
In order to evaluate the performance of different methods on LFs with a higher angular resolution, We used the same training dataset  as the experiment in Sec. \ref{sec:real} but with $5\times5$ uniformly sampled SAIs 
to train another two models  
for $4\times$ and $8\times$ reconstruction, respectively. 
The color augmentation was not applied during training as the color inconsistency issue  does not appear on simulated data.
The rest 19 LF images in the datasets were used for testing.
We converted the LF images to YUV color space, and only used the Y components for training and quantitative evaluation. When generating visual results, the U and V components were up-sampled using bicubic interpolation.

    \begin{figure*}[t]
    \begin{center}
    \includegraphics[width=0.9\linewidth]{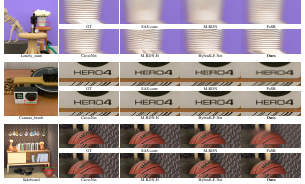}
    \end{center}
    \vspace{-1em}
      \caption{Visual comparisons of different methods on $4\times$ reconstruction from simulated hybrid data. For each algorithm, we provide the zoom-in images of the red block and EPIs constructed at the green line. 
    }
    \label{fig:visual_simulate_4x}
    \end{figure*}
    
    \begin{figure*}[t]
    \begin{center}
    \includegraphics[width=0.9\linewidth]{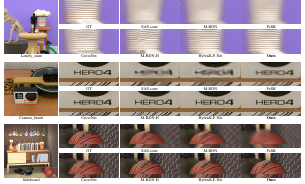}
    \end{center}
    \vspace{-1em}
      \caption{Visual comparisons of different methods on $8\times$ reconstruction from simulated hybrid data. For each algorithm, we provide the zoom-in images of the red block and EPIs constructed at the green line. 
    }
    \label{fig:visual_simulate_8x}
    \end{figure*}

    \begin{table*}[t]
    \centering
    \caption{
    Quantitative comparisons of the average SSIM of EPIs reconstructed by different methods. The best and second best results are colored in \textcolor{red}{ red} and \textcolor{blue}{ blue}, respectively.
    }\vspace{-0.3cm}
    \label{tab:quan_EPI_SSIM}
    \resizebox{0.8\textwidth}{!}{
    \begin{tabular}{c|c c c c c c c|c}
    \toprule[1pt]
    Scale & Bicubic  & SAS-conv \cite{lfssr2018separable}   & M-RDN & PaSR \cite{lfhybrid2014iccp}  & CrossNet \cite{refsr2018crossnet}   & M-RDN-H &  HybridLF-Net \cite{lfhybrid2020fusion}  & Ours\\
    \midrule[0.5pt]
    $4\times$ & 0.819 &	0.900 &	0.891 &	0.916 & 0.957 & 0.973 &	\textcolor{blue}{0.974} & \textcolor{red}{0.976} \\
    $8\times$ & 0.724 & 0.805 &	0.791 &	0.882 &	0.929 & 0.948 &	\textcolor{blue}{0.955} & \textcolor{red}{0.963} \\
    \bottomrule[1pt]
    \end{tabular}
    }
    \end{table*}

\subsubsection{Comparison with state-of-the-art methods}
We compared the proposed approach with state-of-the-art methods for LF reconstruction from the hybrid input, including PaSR \cite{lfhybrid2014iccp}, CrossNet \cite{refsr2018crossnet}, M-RDN-H, and HybridLF-Net \cite{lfhybrid2020fusion}. We also provided comparisons with LF SR methods, i.e., SAS-conv \cite{lfssr2018separable} and M-RDN.
Similar to M-RDN-H, we constructed the baseline model M-RDN by modifying RDN \cite{sisr2018rdn} to adapt to LF data, in which all SAIs of an LF image are stacked along the feature channel and then fed into the residual dense network for spatial SR.
Fig. \ref{fig:mrdnh} shows the network architecture of M-RDN.
Note that all the learning-based methods were re-trained with our training dataset for fair comparisons.

\textbf{Comparison of quantitative results.}
We used PSNR and SSIM to quantitatively measure the quality of the reconstructed HR LF images  from simulated hybrid data
by different methods, and the corresponding results are listed in Table \ref{tab:quan_simulate}, where   
 we can observe that:
\begin{itemize}  
\item the methods with a hybrid input, including PaSR, CrossNet, M-RDN-H, HybridLF-Net, and Ours, significantly outperform those with only an LR LF input, including SAS-conv and M-RDN, which indicates that the extra HR view indeed makes contributions by providing more high-frequency information about the scene, and the five methods for hybrid inputs  have the ability to take advantage of such valuable information to some extent. Also, this observation validates the potential of the hybrid LF imaging; 
\item among methods with a hybrid input, the traditional method PaSR is inferior to others, indicating that a simple model with a small capacity is not enough to model the intricate relations contained in the hybrid input,  while learning-based methods, including CrossNet, M-RDN-H,  HybridLF-Net, and Ours, have much larger capacities; and
\item our approach achieves the highest PSNR/SSIM in average at both scales and exceeds the second best methods (i.e.,  HybridLF-Net \cite{lfhybrid2020fusion}) by around 0.5 dB at $4\times$ and 1 db at $8\times$ reconstruction, demonstrating the great advantage of our method. 
\end{itemize}

\textbf{Comparison of visual results.}
We visually compared different methods for $4\times$ and $8\times$ reconstruction from simulated hybrid data in Figs. \ref{fig:visual_simulate_4x} and \ref{fig:visual_simulate_8x}. 
These results further demonstrate the significant advantages of the proposed approaches over the state-of-the-art ones, i.e., our approach can reconstruct sharper edges and clearer scenes, which are closer to the ground-truth ones. Particularly, for $8\times$ reconstruction, it is very difficult to recover the details without the guidance of an HR view.
From Fig. \ref{fig:visual_simulate_8x}, it can be seen that the patterns in the results of SAS-conv and M-RDN are seriously distorted. 
In contrast, CrossNet, M-RDN-H,  HybridLF-Net, and Ours accept less influence of the scale increasing and can still produce acceptable results. Moreover,  our algorithm successfully preserves the high-frequency details and reconstructs sharper images.

\textbf{Comparison of the LF parallax structure}.
Comparing the 2-D epipolar plane image (EPI) is a straightforward way to evaluate the LF structure qualitatively. 
In the EPI of an LF image, the projections of a single scene point observed in different views construct a straight line. Therefore, we present EPIs constructed from the predictions of different algorithms for comparison.
As shown in Figs. \ref{fig:visual_simulate_4x} and \ref{fig:visual_simulate_8x}, we can observe that the EPIs of our algorithm have clearer line texture and more accurate slops, which demonstrates that our network preserves the LF structure better than others.
Besides, as the ground-truth HR LF images are available in this scenario,  we also evaluated the LF parallax structure of the reconstructed HR LF images 
by different methods both qualitatively.  
Specifically,  
considering that SSIM is a well-known metric to measure the structural similarity between images, we computed the SSIM values over EPIs. 
As listed in Table \ref{tab:quan_EPI_SSIM},  the superiority of our method is demonstrated again, based on the fact that our method produces the highest SSIM values, especially on the $8\times$ reconstruction, which poses great challenges to other methods in preserving the LF parallax structure.

We further  
compared the depth maps estimated from the reconstructed HR LF images by different methods on the simulated data quantitatively. Note that the ground-truth depth map of \textit{Bedroom} is not available, and thus we excluded it for comparison. We still adopted the LF depth estimation algorithm 
in \cite{lfdepth2018jake}  
and used the mean squares error (MSE) between the estimated depth map and its ground-truth to measure the accuracy. 
Table \ref{tab:quan_depth_estimation} lists the quantitative results of each LF image, where it can be seen that the depth maps from the reconstructed LFs by our method achieve the highest and second-highest accuracy on the majority of scenes under the $4\times$ and $8\times$ tasks. 
Besides, the MSE values of our method and other compared methods are even lower than those of GT in some scenes. The reason is that LF depth estimation itself is an open problem, and no method can guarantee perfect estimation, potentially resulting in errors due to the limitation of the estimation method itself. Although the errors introduced by the LF depth estimation method are inevitable, our method achieves top-2 accuracy on the majority of scenes, which is sufficient to demonstrate the advantage of our method.

\begin{table}[t]
    \centering
    \caption{
    Quantitative (MSE) comparisons of the depth estimated from the reconstructed LFs by different methods on simulated hybrid data. The upper and bottom parts show the results of 4$\times$ 
 and 8$\times$ reconstruction, respectively.  
    ``-" indicates that the ground truth depth map of the scene is not available. ``GT" refers to the results of the depth maps estimated from the ground-truth LF data.
    The best and second best results are colored in \textcolor{red}{red} and \textcolor{blue}{blue}, respectively.
    }\vspace{-0.3cm}
    \label{tab:quan_depth_estimation}
    \setlength\tabcolsep{3pt}
    \resizebox{0.48\textwidth}{!}{
    \begin{tabular}{l|c c c c c c |c |c}
    \toprule[1pt]
       LF & \makecell{SAS-conv \\\cite{lfssr2018separable}} &M-RDN &\makecell{PaSR \\\cite{lfhybrid2014iccp}} & \makecell{CrossNet \\\cite{refsr2018crossnet}}    & M-RDN-H     &  \makecell{HybridLF-Net \\\cite{lfhybrid2020fusion}}   &  Ours & GT\\
\midrule[0.5pt]    
       Bedroom             &-&-&-&-&-&-&-&- \\
       Boardgames           &0.060 &0.054 &0.050 &0.053 &\textcolor{blue}{0.044} &0.051 &\textcolor{red}{0.044} &0.045  \\
       Sideboard            &0.137 &0.126 &0.333 &0.130 &0.103 &0.130 &\textcolor{blue}{0.101} &\textcolor{red}{0.098}  \\
       Town                 &0.082 &0.082 &0.122 &0.081 &0.063 &0.081 &\textcolor{blue}{0.063} &\textcolor{red}{0.057}\\
       Antiques             &0.216 &0.190 &0.184 &0.143 &\textcolor{blue}{0.130} &0.143 &\textcolor{red}{0.122} &0.139 \\
       Camera\_brush        &0.111 &0.090 &0.092 &0.063 &0.105 &0.059 &\textcolor{blue}{0.058} &\textcolor{red}{0.039} \\
       Chess                &0.134 &0.126 &\textcolor{blue}{0.042} &0.135 &0.051 &0.047 &0.045 &\textcolor{red}{0.037}  \\
       Coffee\_time         &0.046 &0.048 &0.030 &0.049 &0.045 &0.048 &\textcolor{blue}{0.019} &\textcolor{red}{0.013} \\
       Flowers\_clock       &0.285 &0.152 &0.486 &0.307 &\textcolor{blue}{0.129} &0.332 &0.238 &\textcolor{red}{0.116} \\
       Lonely\_man          &\textcolor{blue}{0.735} &0.950 &1.547 &0.944 &1.058 &0.865 &0.795 &\textcolor{red}{0.702} \\
       Microphone  &0.214 &0.229 &0.430 &0.226 &\textcolor{red}{0.199} &0.210 &\textcolor{blue}{0.204} &0.209  \\
       Pinenuts\_blue       &0.097 &0.117 &0.383 &0.119 &0.112 &0.091 &\textcolor{blue}{0.078} &\textcolor{red}{0.072}  \\
       Rooster\_clock       &0.056 &0.044 &0.052 &0.041 &\textcolor{red}{0.035} &0.042 &\textcolor{blue}{0.039} &0.043 \\
       Roses\_bed           &0.095 &0.113 &0.301 &0.074 &0.091 &\textcolor{blue}{0.074} &0.121 &\textcolor{red}{0.071} \\
       Roses\_table         &\textcolor{red}{1.519} &1.848 &3.292 &\textcolor{blue}{1.555} &1.992 &1.752 &1.933 &2.083 \\
       Toy\_friends         &0.150 &0.145 &0.156 &0.104 &0.095 &0.089 &\textcolor{red}{0.078} &\textcolor{blue}{0.079} \\
       Toys                 &0.124 &0.094 &0.187 &0.123 &0.076 &0.074 &\textcolor{blue}{0.070} &\textcolor{red}{0.054} \\
       Two\_vases           &0.277 &0.234 &0.318 &0.257 &\textcolor{blue}{0.227} &0.355 &0.276 &\textcolor{red}{0.210} \\
       White\_roses         &\textcolor{red}{0.507} &0.824 &2.031 &\textcolor{blue}{0.617} &0.766 &0.634 &0.658 &0.684 \\
      \midrule[0.5pt]
\textbf{Average}    &\textcolor{blue}{0.269} &0.304 &0.558 &0.279 &0.296 &0.280 &0.274 &\textcolor{red}{0.264} \\
\midrule[0.5pt]    
       Bedroom              &-&-&-&-&-&-&-&- \\
       Boardgames           &0.115 &0.078 &0.090 &0.073 &0.077 &0.055 &\textcolor{blue}{0.048} &\textcolor{red}{0.045}  \\
       Sideboard            &0.887 &0.655 &1.178 &0.441 &0.199 &0.157 &\textcolor{blue}{0.123} &\textcolor{red}{0.098}  \\
       Town                 &0.125 &0.113 &0.265 &0.099 &0.089 &0.083 &\textcolor{blue}{0.073} &\textcolor{red}{0.057}\\
       Antiques             &0.423 &0.373 &0.570 &0.202 &0.163 &0.189 &\textcolor{blue}{0.144} &\textcolor{red}{0.139}\\
       Camera\_brush        &0.156 &0.125 &0.276 &0.064 &0.153 &0.102 &\textcolor{blue}{0.055} &\textcolor{red}{0.039}\\
       Chess                &0.128 &0.316 &0.084 &0.256 &0.583 &1.573 &\textcolor{blue}{0.077} &\textcolor{red}{0.037} \\
       Coffee\_time         &0.069 &0.072 &0.091 &\textcolor{blue}{0.029} &0.075 &0.084 &0.030 &\textcolor{red}{0.013}\\
       Flowers\_clock       &0.458 &0.208 &0.490 &0.454 &0.191 &0.197 &\textcolor{blue}{0.170} &\textcolor{red}{0.116}\\
       Lonely\_man          &1.325 &1.236 &2.512 &1.505 &1.459 &1.530 &\textcolor{blue}{1.166} &\textcolor{red}{0.702}\\
       Microphone  &0.319 &0.382 &0.689 &0.301 &0.360 &\textcolor{blue}{0.239} &0.240 &\textcolor{red}{0.209}  \\
       Pinenuts\_blue       &0.172 &0.176 &0.761 &0.301 &0.222 &0.222 &\textcolor{blue}{0.148} &\textcolor{red}{0.072} \\
       Rooster\_clock       &0.108 &0.065 &0.085 &0.051 &0.065 &\textcolor{red}{0.041} &0.046 &\textcolor{blue}{0.043}\\
       Roses\_bed           &0.161 &0.175 &0.547 &0.091 &0.195 &0.106 &\textcolor{blue}{0.081} &\textcolor{red}{0.071}\\
       Roses\_table         &2.346 &2.863 &6.152 &\textcolor{red}{1.940} &2.552 &2.566 &2.244 &\textcolor{blue}{2.083}\\
       Toy\_friends         &0.284 &0.227 &0.313 &0.131 &0.129 &0.139 &\textcolor{blue}{0.085} &\textcolor{red}{0.079}\\
       Toys                 &0.480 &0.401 &0.333 &0.212 &3.070 &0.309 &\textcolor{blue}{0.122} &\textcolor{red}{0.054} \\
       Two\_vases           &0.271 &0.301 &0.431 &0.420 &\textcolor{blue}{0.261} &0.267 &0.776 &\textcolor{red}{0.210}\\
       White\_roses         &0.820 &1.228 &3.076 &\textcolor{blue}{0.776} &0.940 &1.182 &1.357 &\textcolor{red}{0.684}\\
      \midrule[0.5pt]
\textbf{Average}    &0.480 &0.500 &0.997 &0.408 &0.599 &0.502 &\textcolor{blue}{0.388} &\textcolor{red}{0.264} \\
    \bottomrule[1pt]
    \end{tabular}
    }
    \end{table}
    
\subsubsection{Ablation study}
We conducted experiments to demonstrate  that our method is able to leverage multiple LR images and benefit the reconstruction quality of each side view.  
Specifically, we trained two SR models for hybrid inputs with $3\times 3$ and $5\times 5$ LR side views, respectively, both containing a central HR image as a reference. Then we took the super-resolved results of the eight views around the HR central view and compared their average quantitative results of the two SR models in Table \ref{tab:num_side_view}, where we can see that the same eight views can be reconstructed with higher quality when more side views are provided, demonstrating that our model can effectively leverage the \textit{complementary information} of multiple views to improve the reconstruction quality.

\begin{figure*}[t]
\begin{center}
\includegraphics[width=0.9\linewidth]{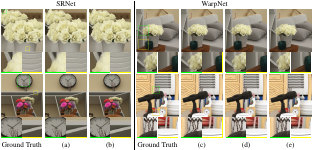}
\end{center}
\vspace{-1.2em}
  \caption{
  Visual comparisons of the intermediate predictions results from SR-Nets and Warp-Nets of HybridLF-Net and our method. (a) HybridLF-Net-S, (b) Ours-S, (c) HybridLF-Net-W, (d) Ours-W (multi-scale), and (e)  Ours-W (single-scale).}
\label{fig:visual_intermediate}
\end{figure*}

Besides, to validate the effectiveness of the newly proposed modifications and architectures in SR-Net and Warp-Net, i.e., using a stack of LR side images in SR-Net, and multi-scale structure in Warp-Net. We qualitatively compared the performance of SR-Net and Warp-Net of HybridLF-Net and our method. We denote the results of SR-Net and Warp-Net in HybridLF-Net as HybridLF-Net-S and HybridLF-Net-W, respectively, and denote the results of SR-Net and Warp-Net in our method as Ours-S and Ours-W (multi-scale), respectively. To directly verify the advantage of the multi-scale Warp-Net over the single-scale one, we also set a baseline, named Ours-W (single-scale), by modifying the Warp-Net in our method with a single-scale structure while leaving other modules unchanged. As shown in Fig. \ref{fig:visual_intermediate}, it can be observed that
\begin{itemize}
    \item Ours-S and Ours-W (multi-scale) can reconstruct sharper edges and clear textures than HybridLF-Net-S and HybridLF-Net-W, respectively, validating the advantage of using a stack of LR side images for SR and multi-scale Warp-Net over those of Hybrid-Net; and
    \item Ours-W (multi-scale) can reconstruct sharp edges at the occlusion boundaries that are closer to the ground truth than Ours-W (single-scale), directly validating the advantage of the multi-scale structure. 
\end{itemize} 
We also quantitatively compared the performance of the above models over the simulated hybrid dataset. As shown in Table \ref{tab:ablation_intermediate}, it can be observed that Ours-S (resp. Ours-W) achieves higher average PSNR and SSIM values than HybridLF-Net-S (resp. HybridLF-Net-W), and Ours-W (multi-scale) produces higher average PSNR and SSIM values than Ours-W (single-scale).

\begin{table}[t]
    \centering
    \caption{Comparison of the reconstruction quality (PSNR/SSIM) on 8$\times$ reconstruction with  3$\times$3 and 5$\times$5 side-views. We refer readers to the supplementary file for the quantitative result of each test LF.
    } \vspace{-0.3cm}
    \label{tab:num_side_view}
    \resizebox{0.6\linewidth}{!}{
    \begin{tabular}{c|c c}
    \toprule
    Scale       &  Ours-$3\times3$ & Ours-5$\times5$\\
    \midrule
    $8\times$ & 38.65/0.983 & \textbf{39.51}/\textbf{0.986} \\
    \bottomrule
    \end{tabular}
    }
\end{table}

\begin{table}[t]
\centering
\caption{Quantitative comparison of the intermediate predictions from SR-Nets and Warp-Nets of HybridLF-Net and our method. The PSNR/SSIM over 19 test LF images on 8$\times$ reconstruction are provided. We refer readers to the supplementary file for the quantitative result of each test LF.}\vspace{-0.3cm}
\label{tab:ablation_intermediate}
\resizebox{0.9\linewidth}{!}{
\begin{tabular}{c|c|c}
\toprule
\multirow{2}{*}{SR-Net}   & HybridLF-Net-S        & 36.02/0.973 \\
                          & Ours-S                & \textbf{37.17/0.978}\\
\midrule
\multirow{3}{*}{Warp-Net} & HybridLF-Net-W        & 33.38/0.959 \\
                          & Ours-W (multi-scale)  & \textbf{33.97/0.963} \\
                          & Ours-W (single-scale) & 33.82/0.961\\
\bottomrule
\end{tabular}
}
\end{table}

\subsubsection{More quantitative analysis}

To investigate the performance of our method with respect to very low-resolution
side views, we re-trained our model on the $16\times$ task, where the resolution of input LR side views 
is $32\times 32$. Additionally, we also provided the results of the $2\times$ task to have
a comprehensive understanding  
of the performance of our method.  
As listed in Table \ref{tab:low_resolution}, we can see that the PSNR/SSIM values indeed decrease
rapidly when the input images have very low resolution but are still within an acceptable (viewable)
range. The possible reason is that the low-resolution side views cannot provide accurate geometric
information to facilitate the propagation of the high-frequency details from the HR central view.

\begin{table}[t]
    \centering
    \caption{Comparison of the reconstruction quality (PSNR/SSIM) of our method on different scales (i.e., side views with various resolutions). The PSNR/SSIM values refer to the average of the simulated hybrid data. We refer readers to the supplementary file for the quantitative result of each test LF.} \vspace{-0.3cm}
    \label{tab:low_resolution}
    \resizebox{\linewidth}{!}{
    \begin{tabular}{c|c c c c}
    \toprule
      Scale       & 2$\times$ & 4$\times$& 8$\times$ & 16$\times$\\
      Size of HR central view & $[512, 512]$ & $[512, 512]$ & $[512, 512]$& $[512, 512]$\\
      Size of LR side views  & $[256, 256]$ &  $[128, 128]$ &  $[64, 64]$ &  $[32, 32]$\\
    \midrule
    PSNR/SSIM      & 43.67/0.993 & 39.91/0.987 & 37.34/0.979 & 32.20/0.944 \\
    \bottomrule
    \end{tabular}
    }
    \end{table}
    
\begin{table}[!h]
    \centering
    \caption{Comparison of the reconstruction quality (PSNR/SSIM) when setting the HR image as the top-left corner or central view. The PSNR/SSIM values refer to the average of the simulated hybrid data. We refer readers to the supplementary file for the result of each test LF.
    } \vspace{-0.3cm}
    \label{tab:corner}
    \resizebox{0.7\linewidth}{!}{
    \begin{tabular}{c|c c}
    \toprule
    Scale       &  Ours-Corner & Ours-Central\\
    \midrule
    $4\times$ & 38.29/0.982  & \textbf{39.91}/\textbf{0.987} \\
    $8\times$ & 35.07/0.966 & \textbf{37.34}/\textbf{0.979} \\
    \bottomrule
    \end{tabular}
    }
\end{table}

\begin{figure}[!t]
\centering
\subfloat[]{\includegraphics[width=0.2\textwidth]{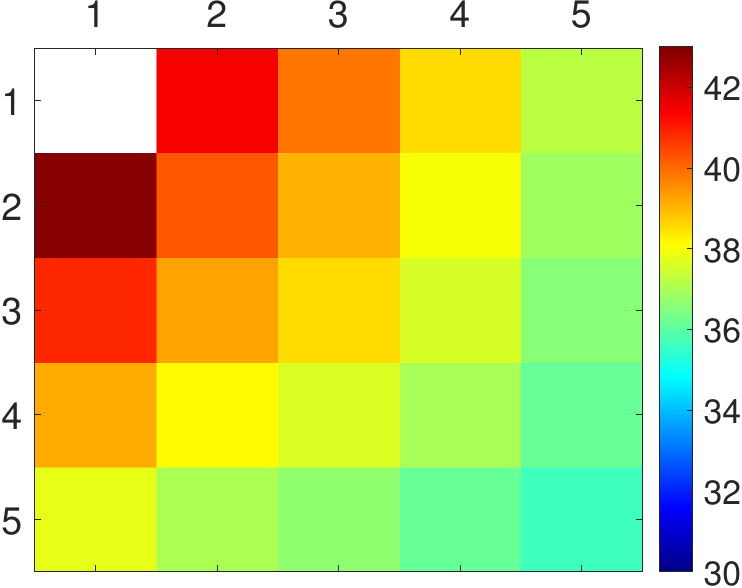}}\hspace{1mm}
\subfloat[]{\includegraphics[width=0.2\textwidth]{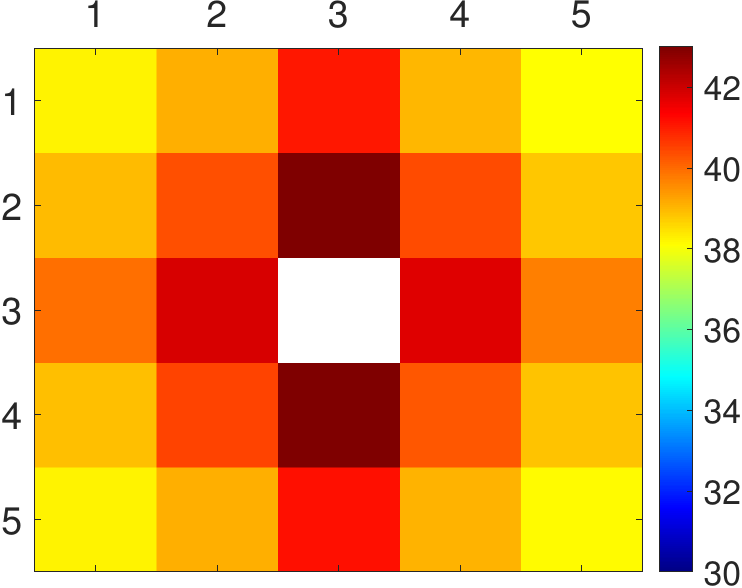}}\hspace{1mm}
\subfloat[]{\includegraphics[width=0.2\textwidth]{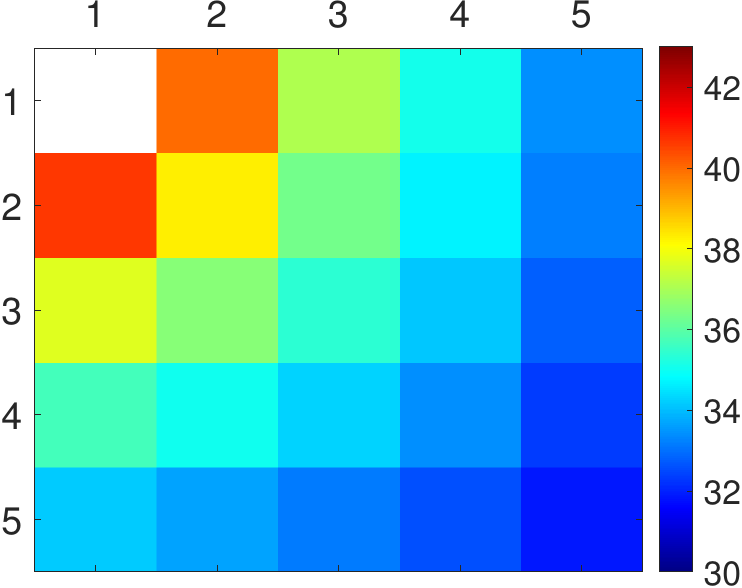}}\hspace{1mm}
\subfloat[]{\includegraphics[width=0.2\textwidth]{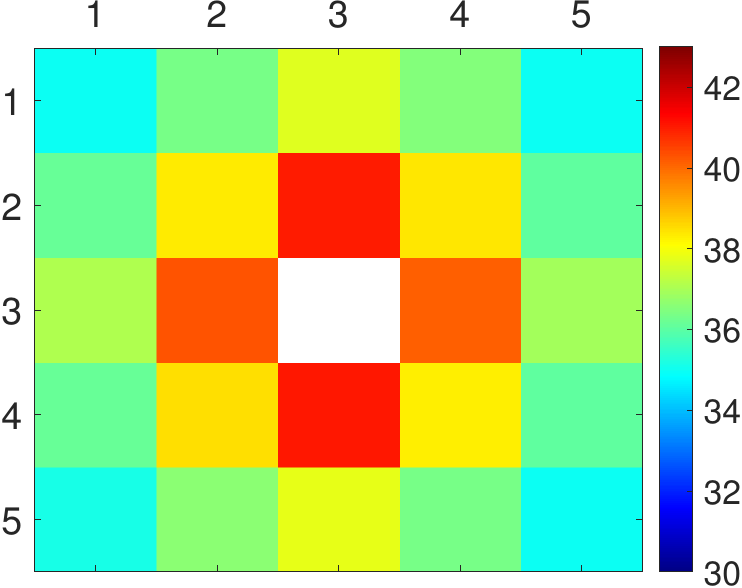}}
\caption{The average PSNR at each angular position of reconstructed LFs by our method under different camera layouts. From left to right: (a) Ours-Corner (4$\times$), (b) Ours-Central (4$\times$), (c) Ours-Corner (8$\times$), (d) Ours-Central (8$\times$). }
\label{fig:corner} 
\end{figure}

Besides, to investigate the generalization capacity of our network in terms of the camera layouts (i.e., putting the HR image at different views), we further conducted experiments on data with the HR view set as the \textbf{top-left} corner view, denoted as \textit{Ours-Corner}. As compared in Table \ref{tab:corner}, we can see that putting the HR image as the central view, denoted as \textit{Ours-Central}, improves the reconstruction quality significantly. Besides, from Fig. \ref{fig:corner}, it can be seen that putting the HR image at the central view balances the quality of all views better.

\begin{figure*}[!t]
\centering
\includegraphics[width=\textwidth]{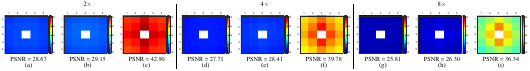}\vspace{-0.3cm}
\caption{Average PSNR at each angular position of reconstructed LFs from different NeRF settings and our method. From left to right: (a) NeRF-LR (2$\times$), (b) NeRF-Hybrid (2$\times$), (c) Ours (2$\times$), (d) NeRF-LR (4$\times$),  (e) NeRF-Hybrid (4$\times$), (f) Ours (4$\times$), (g) NeRF-LR (8$\times$), (h) NeRF-Hybrid (8$\times$), (i) Ours (8$\times$). The average of the PSNR for each setting is shown below each sub-figure. We refer readers to the supplementary file for the quantitative result of each test LF.}
\label{fig:nerf_average} 
\end{figure*}

\subsection{Comparison with NeRF-based Reconstruction}

Recently, the popular view synthesis model NeRF \cite{mildenhall2020nerf} has drawn much attention in the computer vision/graphics community. We also conducted comprehensive experiments to investigate the advantage of our explicit formulation of transferring the information of an HR image to LR ones over the NeRF-based implicit  modeling. 

Generally, we trained an NeRF model, named NeRF-Hybrid, supervised by the hybrid LF image
(i.e., an HR central view surrounded by eight LR side views), to render the LR side views at the
same resolution as the HR central view. It is expected that NeRF-Hybrid could implicitly transfer the
information of the HR central view to the LR side views during training. However, as demonstrated
in a recent work \cite{wang2021nerf}, an NeRF trained with LR images usually generates blurring effects when used
for rendering images with higher resolution. To promote the NeRF trained with LR images to
reconstruct high-quality HR images, \cite{wang2021nerf} proposes a super-sampling strategy, which splits a pixel
of the LR image into multiple sub-pixels and draws a ray for each sub-pixel. Super-sampling
performs supervision by minimizing the loss between the split pixel and the average of the radiances
rendered from its corresponding sub-pixels. Therefore, to generate high-quality HR side views with
only the supervision of their LR counterparts, following \cite{wang2021nerf}, we supervised the side views with the
super-sampling strategy. More specifically, we supervised the side view by calculating the L2 loss
between a typical pixel of the side view and the average of a grid of s $\times$ s radiances rendered from its
corresponding sub-pixels, where s is the scale factor. We supervised the HR central view by
calculating the L2 loss between the input pixel and its rendered radiance. In addition, we trained
NeRF models with all views being LR images, named NeRF-LR, as the baseline to verify the ability
of NeRF-Hybrid in transferring the information of the HR central view to the LR side views.

As the camera parameters of the Inria dataset used in our simulated hybrid LFs  are not available, we conducted experiments in terms of the $2\times$, $4\times$, and $8\times$ reconstruction
tasks only on the HCI dataset including four scenes. The working mechanism of NeRF only allows
us to train NeRF-LR and NeRF-Hybrid, for each scene per task. Fig. \ref{fig:nerf_average} visually shows the average PSNR at each angular position of reconstructed LFs from NeRF-LR, NeRF-Hybrid, and our method, where it can be observed that
\begin{itemize}
    \item  NeRF-Hybrid consistently performs better than NeRF-LR under all scenarios, indicating that
NeRF can implicitly transfer the information of the HR certral view to LR side views; and
    \item  our method achieves much higher PSNR than NeRF-Hybrid in all scenarios, validating the
advantage of explicitly transferring information from the HR certral view to LR side views adopted in our method. Although NeRF-Hybrid regresses a model for each scene separately, it
cannot supervise the LR side views with the ground-truth HR counterparts containing detailed
information (or high-frequency components), resulting in the trained model having a weak
ability to infer this kind of information during rendering HR images, thus limiting performance. This is consistent with the observation of the work \cite{rahaman2019spectral}, i.e., deep networks are biased towards learning lower frequency functions.
\end{itemize}

\section{Conclusion}
\label{sec:con}
We have presented a novel learning-based framework for reconstructing an HR LF image from a hybrid input in an end-to-end fashion. 
The elegant and innovative network architecture enables the proposed framework, a lightweight CNN, to comprehensively exploit  
the underlying properties of the input from two complementary and parallel perspectives.  
Owing to the careful design and the training and data augmentation strategies, our framework trained with simulated hybrid data is able to adapt to real hybrid data by a typical hybrid imaging system very well.
Extensive experimental results demonstrate that our framework  
not only reconstructs HR LF images with higher quality and better LF parallax structure, but also run  
at a relatively high speed, when compared with state-of-the-art approaches.

\section*{Acknowledgment}

We would like to thank the authors of \cite{lfhybrid2017ring} for sharing the datasets captured by their hybrid LF imaging system.

\bibliographystyle{IEEEtran}
\bibliography{IEEEabrv,./reference}

\end{document}